\documentclass[prd,showpacs,floatfix,nofootinbib,twocolumn,amsmath,amssymb]{revtex4}
\usepackage{graphicx,graphics,color,dcolumn,booktabs,bm}
\usepackage{longtable,lscape}
\usepackage{txfonts}
\usepackage{overpic}
\usepackage{amssymb}
\usepackage{indentfirst}
\usepackage{epsfig}
\usepackage{feynmf}   
\usepackage{epstopdf}   
\usepackage{slashed}  
\usepackage{cases}
\usepackage{color}
\usepackage{multirow}

\usepackage[colorlinks, citecolor=blue,anchorcolor=red,menucolor=red, linkcolor=red,filecolor=red,runcolor=red,urlcolor=blue,frenchlinks=red]{hyperref}
\begin{document}

\title{Charmed-strange mesons revisited: mass spectra and strong decays}

\author{Qin-Tao Song$^{1,2,4}$}\email{songqint@impcas.ac.cn}
\author{Dian-Yong Chen$^{1,2}$\footnote{Corresponding author}}\email{chendy@impcas.ac.cn}
\author{Xiang Liu$^{2,3}$\footnote{Corresponding author}}\email{xiangliu@lzu.edu.cn}
\author{Takayuki Matsuki$^{5,6}$}\email{matsuki@tokyo-kasei.ac.jp}
\affiliation{$^1$Nuclear Theory Group, Institute of Modern Physics of CAS,
Lanzhou 730000, China\\CS20141216
$^2$Research Center for Hadron
and CSR Physics, Lanzhou University $\&$ Institute of Modern Physics
of CAS,
Lanzhou 730000, China\\
$^3$School of Physical Science and Technology, Lanzhou University,
Lanzhou 730000, China\\
$^4$University of Chinese Academy of Sciences, Beijing 100049, China\\
$^5$Tokyo Kasei University, 1-18-1 Kaga, Itabashi, Tokyo 173-8602, Japan\\
$^6$Theoretical Research Division, Nishina Center, RIKEN, Saitama 351-0198, Japan}

\begin{abstract}

Inspired by the present experimental status of charmed-strange mesons, we perform a systematic study of the charmed-strange meson family, in which we calculate the mass spectra of the charmed-strange meson family by taking a screening effect into account in the Godfrey-Isgur model and investigate the corresponding strong decays via the quark pair creation model. These phenomenological analyses of charmed-strange mesons not only shed light on the features of the observed charmed-strange states, but also provide important information on future experimental search for the missing higher radial and orbital excitations in the charmed-strange meson family, which will be valuable task in LHCb, forthcoming BelleII and PANDA.

\end{abstract}

 \maketitle

\section{introduction}\label{sec1}
As experiments has largely progressed in the past decade,
more and more charmed-strange states have been reported \cite{Beringer:1900zz}.
Facing the abundant experimental observations, we need to provide an answer, as one crucial task, to the question whether these states can be identified in the charmed-strange meson family, which is not only a valuable research topic relevant to the underlying structure of the newly observed charmed-strange states, but is also helpful to establish the charmed-strange meson family step by step.

It is a suitable time to give a systematic study of the charmed-strange meson family, which includes two main topics, i.e., the mass spectrum and strong decay behavior. At present, we have abundant experimental information of charmed-strange states, which can be combined with the theoretical results to carry out the corresponding phenomenological study.

As the first key step of whole study of the charmed-strange meson family,
the investigation of the mass spectrum of charmed-strange mesons should reflect how a charm quark interacts with a strange antiquark. Godfrey and Isgur proposed the so-called Godfrey-Isgur (GI) model to describe the interaction between $q$ and $\bar{q}$ quarks inside of mesons \cite{Godfrey:1985xj} some thirty years ago. Although the GI model has achieved a great success in reproducing/predicting the low lying mesons,
there exist some difficulties when reproducing the masses of higher radial and orbital excitations, which is because the GI model is a typical quenched model. A typical example of this defect appears in the low mass puzzle of $D_{s0}^\ast(2317)$ \cite{Aubert:2003fg,Besson:2003cp,Abe:2003jk,Aubert:2006bk} and $D_{s1}(2460)$ \cite{Besson:2003cp,Abe:2003jk, Aubert:2003pe, Aubert:2006bk}, where the observed masses of
$D_{s0}^\ast(2317)$ and $D_{s1}(2460)$ are far lower than the corresponding results calculated using the GI model.  A common feature of higher excitations is that they are near the thresholds of meson pairs, which can interact with these higher excitations with the OZI-allowed couplings. Hence, it is unsuitable to use the quenched GI model to describe the mass spectrum of a higher excitated meson and alternatively, we need to adopt an unquenched model.
Although there have been a couple of works studying the heavy-light systems including charmed-strange mesons together with their decay modes \cite{Goity:1998jr,Di_Pierro:2001uu,Bardeen:2003kt, Matsuki:1997da,Matsuki:2007zza,Matsuki:2006rz,Matsuki:2011xp,Ebert:1997nk},
in this work, we would like to modify the GI model such that the screening effect is introduced to {\sl reflect the unquenched pecularity}. In the following section, we present a detailed introduction of the modified GI model.

In this work, we revisit the mass spectrum of the charmed-strange meson to apply the modified GI model , and compare our results with those of the former GI model and experimental data. We would like to see whether this treatment improves the description of the charmed-strange meson spectrum to make the modified GI model reliable. We try further to obtain information of wave functions of charmed-strange mesons, which is important as an input in calculating the decay behavior of the two-body OZI-allowed decay of charmed-strange mesons.

Together with the study of the mass spectrum of charmed-strange mesons,
it is a valuable information of the property of charmed-strange mesons to investigate the decay behavior of charmed-strange mesons. We will adopt the quark pair creation (QPC) model \cite{Micu:1968mk,Le Yaouanc:1972ae,LeYaouanc:1988fx,vanBeveren:1979bd,vanBeveren:1982qb,Bonnaz:2001aj,roberts} to calculate the two-body OZI-allowed decay of charmed-strange mesons, where the corresponding partial and total decay widths are calculated. Through this study, we can further test different possible assignments to the observed charmed-strange states. In addition, we can predict the decay behavior of their partners, which are still missing in experiment. This information is important for experimentalists to further search for these missing charmed-strange mesons, which will be a main task in future experiment.

This work is organized as follows. After Introduction, we will briefly review the research status of the observed charmed-strange states in Sec. \ref{sec2}. In Sec. \ref{sec3}, the GI model is briefly introduced and we give the detailed illustration of how to modify the GI model by introducing the screening effect. The mass spectrum of charmed-strange mesons together with wave fucntions is calculated using the modified GI model. Furthermore, the comparison of our results with experimental data and those from the former GI model is given here. With these preparations, in Sec. \ref{sec4} we further study the two-body OZI-allowed strong decay behaviors of charmed-strange mesons through the QPC model, where the QPC model is briefly introduced. Next, we perform the phenomenological analysis by combining our results with the experimental data. The paper ends with a summary in Sec. \ref{sec5}.

\section{concise review of the observed charmed-strange states}\label{sec2}

\renewcommand{\arraystretch}{1.3}
\begin{table*}[htbp]
\caption{Experimental information of the observed charmed-strange states. \label{table:review} }
\centering
\begin{tabular}{cccccc}
\toprule[1pt] State & Mass (MeV) \cite{Beringer:1900zz} & Width (MeV) \cite{Beringer:1900zz} &
$1^{st}$ observation &Observed decay modes\\ \midrule[1pt]
$D_s$ & $1968.49\pm0.33$ &   & &\\

$D_s^{\ast}$ & $2112.3\pm0.5$& $<1.9$ & &\\

$D_{s0}^\ast(2317)$ & $2317.8\pm0.6$ & $<3.8$  & BaBar \cite{Aubert:2003fg} &$D_s^+ \pi^0$ \cite{Aubert:2003fg}\\
$D_{s1}(2460)$ & $2459.6\pm0.6$ & $<3.5$  & CLEO  \cite{Besson:2003cp}&$D_s^{*+} \pi^0$ \cite{Besson:2003cp}\\
$D_{s1}(2536)$ & $2535.12\pm0.13$  & $0.92\pm0.05$  &ITEP, SERP \cite{Asratian:1987rb}&$D_s^{*+} \gamma$ \cite{Asratian:1987rb}\\
$D_{s2}^\ast(2573)$ & $2571.9\pm0.8$  & $16^{+5}_{-4}\pm3$ \cite{Kubota:1994gn}  &CLEO \cite{Kubota:1994gn}&$D^0 K^+$ \cite{Kubota:1994gn}\\
$D_{sJ}^\ast(2632)$\footnote{Since $D_{sJ}^\ast(2632)$ was only reported by SELEX \cite{Evdokimov:2004iy} and not confirmed by other experiments \cite{Aubert:2004ku}, we do not include this state in our review of this work.} &$2632.5\pm1.7\pm5.0 $ \cite{Evdokimov:2004iy} & $<17$ \cite{Evdokimov:2004iy} & SELEX \cite{Evdokimov:2004iy}&$D^0 K^+$ \cite{Evdokimov:2004iy}\\
$D_{s1}^\ast(2700)$ & $2688\pm4\pm3$ \cite{Aubert:2006mh}  & $112\pm7\pm36$ \cite{Aubert:2006mh}& BaBar \cite{Aubert:2006mh} &$D K$ \cite{Aubert:2006mh}\\
&$ 2708\pm9^{+11}_{-10} $ \cite{Brodzicka:2007aa}&$ 108\pm23^{+36}_{-31} $ \cite{Brodzicka:2007aa}& Belle \cite{Brodzicka:2007aa}&$D^0 K^+$ \cite{Brodzicka:2007aa}\\

&$ 2710\pm2^{+12}_{-7} $ \cite{Aubert:2009ah}&$ 149\pm7^{+39}_{-52} $ \cite{Aubert:2009ah}& BaBar \cite{Aubert:2009ah}&$D^{(*)} K$\cite{Aubert:2009ah} \\
&$ 2709.2\pm1.9\pm4.5 $ \cite{Aaij:2012pc} &$ 115.8\pm7.3\pm12.1 $ \cite{Aaij:2012pc}& LHCb \cite{Aaij:2012pc}&$D K$ \cite{Aaij:2012pc}\\

$D_{sJ}^\ast(2860)$ & $2856.6\pm1.5\pm5.0$ \cite{Aubert:2006mh} & $47\pm7\pm10$ \cite{Aubert:2006mh}& BaBar \cite{Aubert:2006mh}&$D K$ \cite{Aubert:2006mh}\\
&$2862\pm2^{+5}_{-2}$ \cite{Aubert:2009ah}&$48\pm3\pm6$ \cite{Aubert:2009ah}& BaBar \cite{Aubert:2009ah}&$D^{(*)} K$ \cite{Aubert:2009ah}\\
 &$2866.1\pm1.0\pm6.3$ \cite{Aaij:2012pc} & $69.9\pm3.2\pm6.6$ \cite{Aaij:2012pc} & LHCb \cite{Aaij:2012pc}&$D K$ \cite{Aaij:2012pc}\\
$D_{s3}^\ast(2860)$ &$2860.5\pm2.6\pm2.5\pm6.0$ \cite{Aaij:2014xza,Aaij:2014baa}&$53\pm7\pm4\pm6$ \cite{Aaij:2014xza,Aaij:2014baa}&LHCb \cite{Aaij:2014xza,Aaij:2014baa}&$\bar{D}^0K^-$ \cite{Aaij:2014xza,Aaij:2014baa}\\
$D_{s1}^\ast(2860)$ &$2859\pm12\pm6\pm23$ \cite{Aaij:2014xza,Aaij:2014baa}&$159\pm23\pm27\pm72$ \cite{Aaij:2014xza,Aaij:2014baa}&LHCb \cite{Aaij:2014xza,Aaij:2014baa}&$\bar{D}^0K^-$ \cite{Aaij:2014xza,Aaij:2014baa}\\
$D_{sJ}(3040)$ & $3044\pm8^{+30}_{-5}$ \cite{Aubert:2009ah} & $239\pm35^{+46}_{-42}$ \cite{Aubert:2009ah}& BaBar \cite{Aubert:2009ah}&$D^*K$ \cite{Aubert:2009ah}\\\bottomrule[1pt]
\end{tabular}
\end{table*}

In this section, we briefly review the experimental and theoretical status on charmed-strange mesons. First, in Table \ref{table:review} we collect the experimental information of the observed charmed-strange states, which include resonance parameters and the corresponding experiments. Since $D_s^\pm$ and $D_{s}^{*\pm}$ have beeen established to be $1S$ $c\bar{s}$ mesons, our review mainly focuses on possible candidates of higher radial and orbital excitations in the charmed-strange meson family.

\subsection{$D_{s1}(2536)$ and $D_{s2}^\ast(2573)$}
Before 2013, there are only two good candidates for the $1P$ states in the charmed-strange meson family, which are
$D_{s1}(2536)$ and $D_{s2}^\ast(2573)$.

As a charmed-strange meson with $J^P=1^+$,
$D_{s1}(2536)$ was first observed in 1987 by analyzing the $D_s^\ast
\gamma$ invariant mass spectrum of the $\bar{\nu} N$ scattering process
\cite{Asratian:1987rb}, where the measured mass is $(2535 \pm 28)$
MeV. Later, the
ARGUS Collaboration observed this state in the $D^{\ast +} K^0$ final
state, where its mass and width are $M=(2536 \pm 0.6 \pm 2.0)$ MeV
and $\Gamma<4.6$ MeV
\cite{Albrecht:1989yi}l, respectively. Since there does not exist the $D_{s1}(2536)$ signal in the $D^+ K^0$ invariant mass spectrum,
$D_{s1}(2536)$ has an unnatural spin-parity \cite{Albrecht:1989yi}.
In 1993, the CLEO Collaboration measured the ratio of
$\Gamma(D_{s1}(2536) \to D_s^\ast \gamma)$ to $\Gamma(D_{s1}(2536)
\to D^\ast K)$, which is \cite{Alexander:1993nq}
\begin{eqnarray}
\frac{\Gamma(D_{s1}(2536) \to D_s^\ast \gamma)}{ \Gamma(D_{s1}(2536)
\to D^\ast K)} <0.42.
\end{eqnarray}
The $D_{s1}(2536)$ has been confirmed by other groups in different
channels \cite{ Avery:1989ui, Asratian:1993zx, Frabetti:1993vv,
Heister:2001nj, Aubert:2007rva, Abazov:2007wg, Chekanov:2008ac,
Lees:2011um, Belle:2011ad}. The BaBar Collaboration reported this
state in the $D_s^+ \pi^+ \pi^-$ invariant mass spectrum
\cite{Aubert:2006bk}. The Belle Collaboration observed
$D_{s1}(2536) \to D^+ \pi^- K^+$ decay mode \cite{Balagura:2007dya},
where the ratio
$\mathcal{B}(D_{s1}^+(2536) \to D^+ \pi^- K^+)/
\mathcal{B}(D_{s1}^+(2536) \to D^{\ast +}  K^+) =(3.27 \pm 0.18 \pm
0.37) \%$ was obtained. In addition, the measurement of the angular distribution of the $D_{s1}(2536)^+
\to D^{\ast +}K^0_s$ indicates that $D_{s1}(2536)^+
\to D^{\ast +}K^0_s$ dominantly occurs via
$S$-wave with $\Gamma(D_{s1}(2536)^+
\to D^{\ast +}K^0_s)_{S\mathrm{-wave}}/\Gamma(D_{s1}(2536)^+ \to
D^{\ast +}K^0_s)_{\mathrm{total}}=(0.72 \pm 0.05 \pm 0.01)$.
The mass value and narrow width of $D_{s1}(2536)$ are consistent with the theoretical expectation of
the charmed strange meson as $J^P=1^+$ \cite{Godfrey:1986wj}.

In 1994, the CLEO Collaboration observed a charmed-strange meson
$D_{s2}^\ast(2573)$ in the $D^0K^+$ invariant mass
spectrum \cite{Kubota:1994gn}, where the measured resonance
parameters are $M=(2573^{+1.7}_{-1.6} \pm 0.8 \pm 0.5)$ MeV and
$\Gamma =(16^{+5}_{-4} \pm3)$ MeV. In addition, an upper limit of the following ratio was given
\begin{eqnarray}
\frac{\mathcal{B}(D_{s2}^\ast(2573)^{+} \to D^{\ast 0} K^+)}{
\mathcal{B}(D_{s2}^\ast(2573)^{+} \to D^{ 0} K^+) } <0.33,
\end{eqnarray}
which is from the search for the decay mode $D_{s2}^\ast(2573)^+ \to
D^{\ast 0} K^+$ \cite{Kubota:1994gn}.
At present, $D_{s2}^\ast(2573)$ is a good candidate of a charmed-strange meson with $1^3P_2$ since its mass is consistent with the theoretical prediction \cite{Godfrey:1986wj}. The $D_{s2}^\ast(2573)$ had
subsequently been confirmed by some other collaborations in the $D^0K^+$
invariant mass spectrum \cite{ Aaij:2011ju,Aubert:2006mh,
Evdokimov:2004iy,Heister:2001nj,Albrecht:1995qx}.

\subsection{$D_{s0}^\ast(2317)$ and $D_{s1}(2460)$}

In 2003, the BaBar Collaboration observed a new charmed-strange state
near 2.32 MeV in the $D_s^+ \pi^0$ invariant mass distribution of the $B$ decay, which is
named as $D_{s0}^\ast(2317)$ \cite{Aubert:2003fg}. Later, the CLEO
Collaboration confirmed this state in the $D_s^+ \pi^0$ channel and
reported another narrow charmed-strange state $D_{s1}(2460)$ \cite{Besson:2003cp}.
In addition, the ratio
\begin{eqnarray}
\frac{\mathcal{B}(D_{s0}^\ast(2317)^+ \to D_s^{\ast +}
\gamma)}{\mathcal{B}(D_{s0}^\ast(2317)^+ \to D_s^{ +} \pi^0)} <0.059
\end{eqnarray}
was obtained in Ref. \cite{Besson:2003cp} while the Belle and BaBar
collaborations gave the upper bound of this ratio as 0.18 and
0.16, respectively \cite{Abe:2003jk,Aubert:2006bk}.
The $D_{s1}(2460)$ was also confirmed by the Belle and BaBar collaborations \cite{Abe:2003jk,Aubert:2003pe,Aubert:2006bk}.

If assigning $D_{s0}^\ast(2317)$ and $D_{s1}(2460)$ as charmed-strange mesons with quantum numbers $J^P=0^+$ and $J^P=1^+$, respectively, there exists the low mass puzzle for $D_{s0}^\ast(2317)$ and $D_{s1}(2460)$, i.e.,
the theoretical masses of charmed-strange mesons with
$0^+$ and $1^+$ are 2.48 GeV and 2.53 GeV, respectively, predicted by the old but successful model proposed in Refs. \cite{Godfrey:1986wj,Godfrey:1985xj}, which values are far larger than the corresponding experimental data. These peculiarities have also stimulated theorists' extensive interest in exploring their inner structures and the exotic state explanations to $D_{s0}^\ast(2317)$ and $D_{s1}(2460)$ were especially proposed in Refs. \cite{Barnes:2003dj,Cheng:2003kg,Chen:2004dy}.

Since $D_{s0}^\ast(2317)$ and $D_{s1}(2460)$ are near and below the thresholds of $DK$ and $D^*K$, respectively,
the coupled-channel effect, which is an important non-perturbative QCD effect, should be considered in understanding  the low mass puzzle. This means that $D_{s0}^\ast(2317)$ and $D_{s1}(2460)$ are still categorized as the conventional charmed-strange meson family \cite{vanBeveren:2003kd,Dai:2006uz,Liu:2009uz}.

\subsection{  $D_{s1}^\ast(2700)$}

In 2006, the BaBar Collaboration observed a broad structure in the $DK$ invariant mass spectrum, which was named as $D_{s1}^\ast(2700)$ \cite{Aubert:2006mh} and the mass and width are $M=(2688 \pm 4 \pm3)$ MeV and $\Gamma=(112\pm 7 \pm 36)$ MeV, respectively. Later, $D_{s1}^\ast(2700)$ was confirmed by the Belle Collaboration in the $DK$ invariant mass spectrum of $B \to \bar{D}^0 \{D^0 K^+\} $ process \cite{Brodzicka:2007aa}. The angular momentum and parity of $D_{s1}^\ast(2700)$ are determined to be $J=1$ and $P=-$ by the helicity angle distribution and by its decay to two pseudoscalar mesons, respectively. The Babar Collaboration reported the $D^\ast K$ decay mode of $D_{s1}(2700)$ in Ref. \cite{Aubert:2009ah}, and obtained a ratio of $\mathcal{B}(D_{s1}^\ast(2710) \to D^\ast K)$ to $\mathcal{B}(D_{s1}^\ast(2710) \to DK)$ as \cite{Aubert:2009ah}
\begin{eqnarray}
\frac{\mathcal{B}(D_{s1}^\ast(2710) \to D^\ast K)}{\mathcal{B}(D_{s1}^\ast(2710) \to DK)} = 0.91 \pm 0.13 \pm 0.12.
\label{r1}
\end{eqnarray}
In 2012, the LHCb Collaboration also observed $D_{s1}^\ast(2700)$ in the $DK$ mass spectrum \cite{Aaij:2012pc}.

As a vector charmed-strange state, the measured mass of $D_{s1}^\ast(2700)$ is close to the prediction of the $2^3S_1$ charmed-strange meson \cite{Godfrey:1985xj}. The strong decay behavior of a $c\bar{s}$ state of $2^3S_1$ is investigated using the QPC model in Ref. \cite{Zhang:2006yj}. The ratio $\mathcal{B}(D_{s1}^\ast(2700) \to D^\ast K)/\mathcal{B}(D_{s1}^\ast(2700) \to DK)$ evaluated by the effective Lagrangian approach, however, favors the $2^1S_0$ assignment for $D_{s1}^\ast(2700)$ \cite{Colangelo:2007ds}. Besides the decay behavior, the production of $D_{s1}^\ast(2700)$ from the $B$ meson decay was calculated by a naive factorization method, which shows that $D_{s1}^\ast(2700)$ could be explained as the first radial excitation of $D_{s}^\ast(2112)$ \cite{Wang:2009as}.

In Refs. \cite{Close:2006gr, Li:2009qu}, $D_{s1}^\ast(2700)$ was assigned as a mixing of $2^3S_1$ and $1^3D_1$ $c\bar{s}$ state, and this assignment can be supported by the study of the decay behavior obtained by the QPC model. The obtained ratio of $\mathcal{B}(D_{s1}^\ast(2710) \to D^\ast K)/\mathcal{B}(D_{s1}^\ast(2710) \to DK)$ is also consistent with the experimental measurement \cite{Li:2009qu} from BaBar \cite{Aubert:2009ah}. In addition,  evaluation by the constituent quark model \cite{Zhong:2009sk} and the Regge Phenomenology \cite{Li:2007px} also support the assignment of $D_{s1}^\ast(2710)$ as a mixing of the $2^3S_1$ and $1^3D_1$ $c\bar{s}$ states. Other than a standard interpretation of $D_{s1}^\ast(2700)$ as a $c\bar{s}$ state, a molecular state explanation
was proposed in Ref. \cite{Vinodkumar:2008zd}, which is based on a potential model.

\subsection{$D_{sJ}^\ast(2860)$, $D_{s1}^\ast(2860)$ and $D_{s3}^\ast(2860)$}
The $D_{sJ}^\ast(2860)$ was first discovered by the BaBar Collaboration in the $DK$ invariant mass spectrum of the inclusive process $e^+e^- \to DK X$ \cite{Aubert:2006mh}. Its resonance parameters are reported as $M=(2856.6 \pm 1.5\pm 5.0)$ MeV and $\Gamma=(48 \pm 7\pm10)$ MeV. This state was confirmed in the $D^\ast K $ mode by the BaBar Collaboration \cite{Aubert:2009ah} again and the ratio $\mathcal{B}(D_{sJ}^\ast(2860)^+ \to D^\ast K )/\mathcal{B}(D_{sJ}^\ast(2860)^+ \to D K )$ was measured as
\begin{eqnarray}
\frac{\mathcal{B}(D_{sJ}^\ast(2860)^+ \to D^\ast K )}{\mathcal{B}(D_{sJ}^\ast(2860)^+ \to D K )} = 1.10 \pm 0.15 \pm 0.19.\label{a1}
\end{eqnarray}

Very recently, the LHCb Collaboration reported their new measurement of the structure around 2.86 GeV in the $\bar{D}^0 K^-$ invariant mass distribution of $B^0_s \to \bar{D}^0K^- \pi^+$ \cite{Aaij:2014xza,Aaij:2014baa}. The amplitude analysis of this decay indicates that the structure at $m_{\bar{D}^0K^-} \simeq 2.86$ GeV contains both $J^P=1^-$ and $3^-$ components corresponding to $D_{s1}^\ast(2860)$ and $D_{s3}^\ast(2860)$, respectively, where the resonance parameters of $D_{s1}^\ast(2860)$ and $D_{s3}^\ast(2860)$ are
\begin{eqnarray}
M_{D_{s1}^\ast(2860)}&=&(2859 \pm 12\pm 6 \pm 23\ )\,\mathrm{MeV},\\
\Gamma_{D_{s1}^\ast(2860)}&=&(159 \pm 23\pm 27 \pm 72\ ) \,\mathrm{MeV},\\
M_{D_{s3}^\ast(2860)}&=&(2860.5 \pm 2.6\pm 2.5 \pm 6.0\ )\,\mathrm{MeV},\\
\Gamma_{D_{s3}^\ast(2860)}&=&(53 \pm 7\pm 4 \pm 6\ )\,\mathrm{MeV}.
\end{eqnarray}

Before the measurement by the LHCb Collaboration, the properties of $D_{sJ}^\ast(2860)$ have been widely discussed. The possibility of $D_{sJ}^\ast(2860)$ as the first radial excitation of $D_{s0}^\ast(2317)$ has been ruled out due to the observation of $D_{sJ}^\ast(2860) \to D^\ast K$ \cite{Aubert:2009ah}. According to the calculation by the QPC model \cite{Zhang:2006yj}, the Regge phenomenology \cite{Li:2007px}, chiral quark model \cite{Zhong:2008kd}, and the flux tube model \cite{Chen:2009zt}, $D_{sJ}^\ast(2860)$ can be assigned to a $1^3D_3$ charmed strange meson. However, different approaches give different ratios of $\mathcal{B}(D_{sJ}^\ast(2860) \to D^\ast K) /\mathcal{B}(D_{sJ}^\ast(2860) \to DK)$. For example, this ratio was estimated to be 0.36 by the effective Lagrangian method \cite{Colangelo:2007ds} and to be 0.8 by the QPC model \cite{Li:2009qu}. At present calculation, the $J^P=3^-$ assignment to $D_{sJ}^\ast(2860)$ is still possible.

In Ref. \cite{Li:2009qu}, the $2S$-$1D$ mixing was proposed to explain $D_{sJ}^\ast(2860)$, where $D_{sJ}^\ast(2860)$ and $D_{s1}^\ast(2710)$ are treated as a mixture of $2^3S_1$ and $1^3D_1$ states in the charmed-strange meson family. With a proper mixing angle, the ratios of $\mathcal{B}(D_{sJ}^\ast(2860) \to D^\ast K) /\mathcal{B}(D_{sJ}^\ast(2860) \to DK)$ and $\mathcal{B}(D_{sJ}^\ast(2700) \to D^\ast K) /\mathcal{B}(D_{sJ}^\ast(2700)  \to DK)$ can be well explained simultaneously.
In Ref. \cite{ Zhong:2009sk} the authors proposed a two-state scenario for $D_{sJ}^\ast(2860)$, one resonance is likely to be the $1^3D_3$ and the other resonance seems to be the higher mixing state of $1^1D_2-1^1D_2$.
Although the $J^P=0^+$ assignment to $D_{sJ}^\ast(2860)$ has been ruled out by the experimental measurement of $D_{sJ}^\ast(2860) \to D^\ast K$, the authors of Ref. \cite{vanBeveren:2009jq} indicated that there exist two resonances around $2.86$ GeV with $J^P=0^+$ and $J^P=2^+$. In the $DK$ invariant mass spectrum, the structure near 2.86 GeV should contain these two resonances while in the $D^\ast K$ invariant mass spectrum, only one resonance of $J^P=2^+$ should be included.

Following the measurement by the LHCb Collaboration \cite{Aaij:2014xza,Aaij:2014baa}, the decay behaviors of $D_{s1}^\ast(2860)$ and $D_{s3}^\ast(2860)$ were evaluated by the QPC model \cite{Song:2014mha, Godfrey:2014fga},
which show that  these states can be good candidates of the $1D$ states in the charmed-strange meson family. By using the QCD Sum Rule, the masses of $1D$ charmed-strange mesons were calculated in Ref. \cite{Zhou:2014ytp}, which also supports the explanation of the $1D$ charmed-strange mesons. In addition, another study of the decay behaviors of theses states were performed in Ref. \cite{Wang:2014jua}, where the effective Lagrangian approach was adopted.

Before closing the review of $D_{sJ}^\ast(2860)$, $D_{s1}^\ast(2860)$ and $D_{s3}^\ast(2860)$, we need to comment on the measurement of the ratio in Eq. (\ref{a1}). Since the new measurement by LHCb  \cite{Aaij:2014xza,Aaij:2014baa} is given, thid ratio must be changed according to which state is assigned to $D_{sJ}^\ast$ in both denomenator and numerator. Thus, we do not suggest to adopt the old data of this ratio in Eq.~(\ref{a1}) when checking the theoretical result. We also expect new measurement of this ratio when considering the LHCb results \cite{Aaij:2014xza,Aaij:2014baa}, which will be helpful to pin down the different explanations.

\subsection{$D_{sJ}(3040)$}

Besides confirming $D_{s1}^\ast(2700)$ and $D_{sJ}^\ast(2860)$ in the $D^\ast K$ invariant mass distribution, the BaBar Collaboration also observed a new broad structure with mass $M=(3044 \pm 8^{+30}_{-5})$ MeV and width $\Gamma=(239 \pm 35^{+46}_{-42})$ MeV. The negative result of a decay $D_{sJ}(3040) \to DK$ suggests unnatural parity for $D_{sJ}(3040)$.

The observed mass of $D_{sJ}(3040)$ and its unnatural parity are consistent with the quark model prediction for $2^3P_1$ charmed strange meson \cite{Godfrey:1985xj}, which is the first radial excitation of $D_{s1}(2460)$. The calculations in the QPC model also support that $D_{sJ}(3040)$ can be categorized as $1^+$ state in a $(0^+,1^+)$ spin doublet \cite{Sun:2009tg} of the heavy quark symmetry. In addition, the calculations in the flux tube model \cite{Chen:2009zt}, the constitute quark model \cite{Zhong:2009sk,Xiao:2014ura} and the effective approach \cite{Colangelo:2010te} also indicate that a possibility of a $D_{sJ}(3040)$ as a $1^+$ charmed-strange meson can not be ruled out. References \cite{Segovia:2012cd, Segovia:2013wma} calculated the decay width of $D_{sJ}(3040)$ as $n(J^P)=3(1^+)$ or $4(1^+)$ state, which is rather large but still compatible with the experimental data. Besides $J^P=1^+$ assignment to $D_{sJ}(3040)$, the possibility that $D_{sJ}(3040)$ as a mixture of $1^3D_2$ and $1^1D_2$ charmed-strange mesons was discussed with an effective Lagrangian approach \cite{Colangelo:2010te}.

As can be seen in the above review of status of the observed charmed-strange states,
we notice that more and more candidates of higher radial and orbital charmed-strange mesons were reported in the past decade. It is a good opportunity to carry out the systematic study of charmed-strange mesons now, which will deepen our understanding of the charmed-strange meson family and will provide more abundant information for experimentalists to further search for higher radial and orbital charmed-strange mesons in future experiment.

In the following sections, we start to investigate charmed-strange mesons on their mass spectrum and two-body OZI-allowed decay behaviors.

\section{Mass spectra}\label{sec3}

In this work, we employ the modified relativistic quark model to calculate the mass spectra and wave functions of charmed-strange mesons because owing to the peculiarity of charmed-strange mesons, relativistic effects and unquenched effects cannot be ignored. In 1985, Godfrey and Isgur proposed the GI model to describe the meson spectra with a great success, especially for the low lying mesons \cite{Godfrey:1985xj} and later Godfrey and Kokoski studied $P$-wave heavy-light systems \cite{Godfrey:1986wj}. Since a coupled-channel effect becomes more important for higher radial and orbital excitations, we need to include this effect in calculating the mass spectrum, which motivates us to modify the GI model by introducing the screened potential, which is partly equivalent description to the coupled-channel effect \cite{Li:2009zu}.

In the following, we first give a brief review of the GI model and then present how to introduce the screened potential.

\subsection{Brief review of the Godfrey-Isgur model}

The interaction between quark and antiquark of the GI model \cite{Godfrey:1985xj} is described by the Hamiltonian
\begin{eqnarray}
\tilde{H}=\left(p^2+m_1^2\right)^{1/2}+ \left(p^2+m_2^2\right)^{1/2} +V_{\mathrm{eff}}\left(\textbf{p},\textbf{r}\right), \label{Eq:Htot}
\end{eqnarray}
where $V_{\mathrm{eff}}(\textbf{p},\textbf{r}) = \tilde{H}^{\mathrm{conf}} +\tilde{H}^{\mathrm{hyp}} +\tilde{H}^{\mathrm{SO}}$ is the effective potential of the $q\bar{q}$ interaction.
$V_{\mathrm{eff}}(\textbf{p},\textbf{r})$ contains two main ingredients. The first one is a short-distance $\gamma^{\mu}\otimes \gamma_{\mu}$ interaction of one-gluon-exchange and the second is a long-distance $1\otimes1$ linear confining interaction which is at first employed byt the Cornell group and is suggested by the lattice QCD. This effective potential can be obtained by on-shell $q\bar{q}$ scattering amplitudes in the center-of-mass (CM) frame \cite{Godfrey:1985xj}.

In the nonrelativistic limit,
$V_{\mathrm{eff}}(\textbf{p},\textbf{r})$ is tansformed into the standard nonrelativistic potential
$V_{\mathrm{eff}}(r)$.
\begin{eqnarray}
V_{\mathrm{eff}}(r)=H^{\mathrm{conf}}+H^{\mathrm{hyp}}+H^{\mathrm{SO}}\label{1}
\end{eqnarray}
with

\begin{eqnarray}
H^{\mathrm{conf}}=c+br+\frac{\alpha_s(r)}{r}\bm{F}_1\cdot\bm{F}_2, \label{3}
\end{eqnarray}

where $H^{\mathrm{conf}}$ is the spin-independent potential and contains a constant term, a linear confining
potential and a one-gluon exchange potential. The subscripts 1 and 2 denote quark
and antiquark, respectively. The second term of r.h.s. of Eq. (\ref{1}) is  the color-hyperfine interaction, i.e.,
\begin{eqnarray}
H^{\mathrm{hyp}}&=&-\frac{\alpha_s(r)}{m_1m_2}\Bigg[\frac{8\pi}{3}\bm{S}_1\cdot\bm{S}_2\delta^3 (\bm r)+\frac{1}{r^3}\Big(\frac{3\bm{S}_1\cdot\bm r \bm{S}_2\cdot\bm r}{r^2}\nonumber\\
&& -\bm{S}_1\cdot\bm{S}_2\Big)\Bigg] \bm{F}_1\cdot\bm{F}_2.
\end{eqnarray}
The third term of r.h.s. of Eq. (\ref{1}) is the spin-orbit interaction,
\begin{eqnarray}
H^{\mathrm{SO}}=H^{\mathrm{SO(cm)}}+H^{\mathrm{SO(tp)}}.
\end{eqnarray}
Here, $H^{\mathrm{SO(cm)}}$ is the
color-magnetic term and $H^{\mathrm{SO(tp)}}$ is the Thomas-precession term, i.e.,
\begin{eqnarray}
H^{\mathrm{SO(cm)}}=-\frac{\alpha_s(r)}{r^3}\left(\frac{1}{m_1}+\frac{1}{m_2}\right)\left(\frac{\bm{S}_1}{m_1}+\frac{\bm{S}_2}{m_2}\right)\cdot
\bm{L}(\bm{F}_1\cdot\bm{F}_2),
\end{eqnarray}
\begin{eqnarray}
H^{\mathrm{SO(tp)}}=\frac{-1}{2r}\frac{\partial H^{\mathrm{conf}}}{\partial
r}\Bigg(\frac{\bm{S}_1}{m^2_1}+\frac{\bm{S}_2}{m^2_2}\Bigg)\cdot \bm{L}.
\end{eqnarray}
In the above expressions, $\bm{S}_1/\bm{S}_2$ denotes the spin of the
quark/antiquark and $\bm{L}$ is the orbital momentum between
quark and antiquark. $\bm{F}$ is related to the Gell-Mann matrix by
$\bm{F}_1=\bm{\lambda}_1/2$ and $\bm{F}_2=-\bm{\lambda}^*_2/2$. For a
meson, $\langle\bm{F}_1\cdot\bm{F}_2\rangle=-4/3$.
%
%

The GI model constructed by Godfrey and Isgur is a
relativized quark model, where relativistic effects are embeded into the model mainly in two ways.

Firstly, a smearing function $\rho_{12}(\bm{r-r}')$ is introduced to incorporate effects of an internal motion inside a hadron and nonlocality of interactions between quark and antiquark. A general form is given by
%
%
\begin{eqnarray}
&&\tilde{f}(r)=\int d^3 \bm{r}' \rho_{12}(\bm{r}-\bm{r}') f(r') \label{4}
\end{eqnarray}
with
\begin{eqnarray}
&&\rho_{12}(\bm{r-r}')=\frac{\sigma^3_{12}}{\pi^{3/2}}\text{exp}\left(-\sigma^2_{12}(\bm{r-r}')^2\right), \label{4-1} \\
&&\sigma_{12}^2=\sigma_0^2\Bigg[\frac{1}{2}+\frac{1}{2}\left(\frac{4m_1m_2}{(m_1+m_2)^2}\right)^4\Bigg]+
s^2\left(\frac{2m_1m_2}{m_1+m_2}\right)^2, \nonumber
\end{eqnarray}
where {$\sigma_0=1.80$ GeV and $s=1.55$ are the universal parameters in the GI model,} 
$m_1$ and $m_2$ are masses of quark and antiquark, respectively.
If $m_1=m_2$ and both of them become large, $\sigma_{12}$ also becomes large and $\rho_{12}(\bm{r-r}')\to\delta^3(\bm{r-r}')$ and hence this smearing is effective for a heavy quarkonium because the second term becomes $(sm_1)^2$.
If $m_2 >> m_1$, $\sigma_{12}$ gets its minimum and suitable for describing a heavy-light system like the charmed-strange mesons.
%

%
%
Secondly, {a general formula of the potential should depend on the CM momentum of the interacting quarks. This effect is taken into account by introducing momentum-dependent factors in the interactions and the factors will go to unity in the nonrelativistic limit. In a semiquantitatively relativistic treatment, the smeared Coulomb term $\tilde{G}(r)$ and the smeared hyperfine interactions $\tilde{V}_i$ should be modified according to,
\begin{eqnarray}
\tilde{G}(r) &\to& \left(1+\frac{p^2}{E_1 E_2}\right)^{1/2} \tilde{G}(r) \left( 1+\frac{p^2}{ E_1 E_2}\right)^{1/2},\nonumber\\
\frac{\tilde{V}_i(r)}{m_1 m_2} &\to& \left( \frac{m_1 m_2}{E_1 E_2}\right)^{1/2+\epsilon_i}  \frac{\tilde{V}_i(r)}{m_1 m_2} \left( \frac{m_1 m_2}{E_1 E_2}\right)^{1/2+\epsilon_i},\label{eqGV}
\end{eqnarray}
where $E_1$ and $E_2$ are the energes of the quark and antiquark in the meson, $\epsilon_i$'s are expected to be small numbers for a different type of hyperfine interactions, i.e., the contract, tensor, vector spin-orbit and scalar spin-orbit potentials, into which $\tilde V_i(r)$ are classified. The particular values of $\epsilon_i$ and other parameters are given in Table II of Ref.\cite{Godfrey:1985xj}.
}

%
%

Diagonalizing the Hamiltonian given by Eq.~(\ref{Eq:Htot}) by a simple
harmonic oscillator (SHO) basis and using a variational method, we get the mass spectrum and wave
function of a meson. More details of the GI model can be found in Appendices of Ref. \cite{Godfrey:1985xj}.

\subsection{Modified GI model with the screening effect}

Although the GI model has achieved a great success in describing the meson spectrum \cite{Godfrey:1985xj}, there still exists discrepancy between the predictions given by the GI model and recent new experimental observations. A typical example is that the masses of the observed $D_{s0}(2317)$ \cite{Aubert:2003fg,Besson:2003cp,Abe:2003jk,Aubert:2006bk}, $D_{s1}(2460)$ \cite{Besson:2003cp,Abe:2003jk, Aubert:2003pe, Aubert:2006bk} and $X(3872)$ \cite{Choi:2003ue} deviate from the corresponding values expected by the GI model \cite{Godfrey:1985xj}. Later, theorists realized that these discrepancies are partly caused by coupled channel effects \cite{Eichten:2004uh,vanBeveren:2003kd,Dai:2006uz,Liu:2009uz}. Furthermore,
another remedy can be made to adjust masses by screening the color charges at distances greater than about 1 fm \cite{Born:1989iv} which spontaneously creates light quark-antiquark pairs.
%
%
The screening effect has been confirmed by unquenched Lattice QCD and some holographic models \cite{Bali:2005fu,Armoni:2008jy,Bigazzi:2008gd,Namekawa:2011wt}.

There were several progresses on the study of the meson mass spectrum by considering the screening effect \cite{Li:2009ad,Mezoir:2008vx}. In Ref. \cite{Li:2009ad}, the authors adopted the screened potential to compute the charmonium spectrum \cite{Li:2009zu}. Mezzoir {\it et al.} in \cite{Mezoir:2008vx} carried out the investigation of highly excited light
unflavored mesons by flattening the linear potential $br$ above a certain
saturation distance $r_s$. However, the study of a screening effect for a heavy-light meson system is still absent at present, which is the reason why we introduce the screening effect into the GI model in this work.

In order to take into account the screening effect, we need to make a replacement in Eq. (\ref{3})   \cite{Chao:1992et,Ding:1993uy} as
\begin{eqnarray}
br\to V^{\mathrm{scr}}(r)=\frac{b(1-e^{-\mu r})}{\mu},
\end{eqnarray}
where $V^{\mathrm{scr}}(r)$ behaves like a linear potential $br$ at short distances and approaches to $\frac{b}{\mu}$ at long distances.
Then, we further modify $V^{\mathrm{scr}}(r)$ as the way given in Eq. (\ref{4}),
\begin{eqnarray}
\tilde V^{\mathrm{scr}}(r)&=& \int d^3 \bm{r}^\prime
\rho_{12}(\bm{r-r^\prime})\frac{b(1-e^{-\mu r'})}{\mu}.\label{5}
\end{eqnarray}
By inserting Eq.~(\ref{4-1}) into the above expression, the concrete expression for $\tilde V^{\mathrm{scr}}(r)$ is given by
\begin{eqnarray}
\tilde V^{\mathrm{scr}}(r)&=& \frac{b}{\mu r}\Bigg[r+e^{\frac{\mu^2}{4 \sigma^2}+\mu r}\frac{\mu+2r\sigma^2}{2\sigma^2}\Bigg(\frac{1}{\sqrt{\pi}}
\int_0^{\frac{\mu+2r\sigma^2}{2\sigma}}e^{-x^2}dx-\frac{1}{2}\Bigg) \nonumber\\
&&-e^{\frac{\mu^2}{4 \sigma^2}-\mu r}\frac{\mu-2r\sigma^2}{2\sigma^2}\Bigg(\frac{1}{\sqrt{\pi}}
\int_0^{\frac{\mu-2r\sigma^2}{2\sigma}}e^{-x^2}dx-\frac{1}{2}\Bigg)\Bigg] \label{Eq:pot}
\end{eqnarray}
with
%
%
\begin{eqnarray}
\frac{1}{r}\frac{\partial}{\partial r} \tilde V^{\mathrm{scr}}(r)&=&\frac{b}{\mu r^2}\Bigg[e^{\frac{\mu^2}{4 \sigma^2}+\mu r}\frac{\mu^2 r+2r^2\sigma^2\mu-\mu}{2\sigma^2 r}\nonumber\\
&&\times\Bigg(\frac{1}{\sqrt{\pi}}\int_0^{\frac{\mu+2r\sigma^2}{2\sigma}}e^{-x^2}dx-\frac{1}{2}\Bigg)+
e^{\frac{\mu^2}{4 \sigma^2}-\mu r} \nonumber\\
 &&\times\frac{\mu^2 r-2r^2\sigma^2\mu+\mu}{2\sigma^2 r}\Bigg(\frac{1}{\sqrt{\pi}}
\int_0^{\frac{\mu-2r\sigma^2}{2\sigma}}e^{-x^2}dx \nonumber\\
&&-\frac{1}{2}\Bigg)+\frac{\mu}{\sqrt{\pi}\sigma}e^{-\sigma^2r^2}\Bigg],
\end{eqnarray}
where $\tilde V(r)$ is given in the footnote in this page and $\sigma=\sigma_{12}$ given by Eq.~(\ref{4-1}).

\begin{figure}[htpb]
\includegraphics[width=0.49\textwidth]{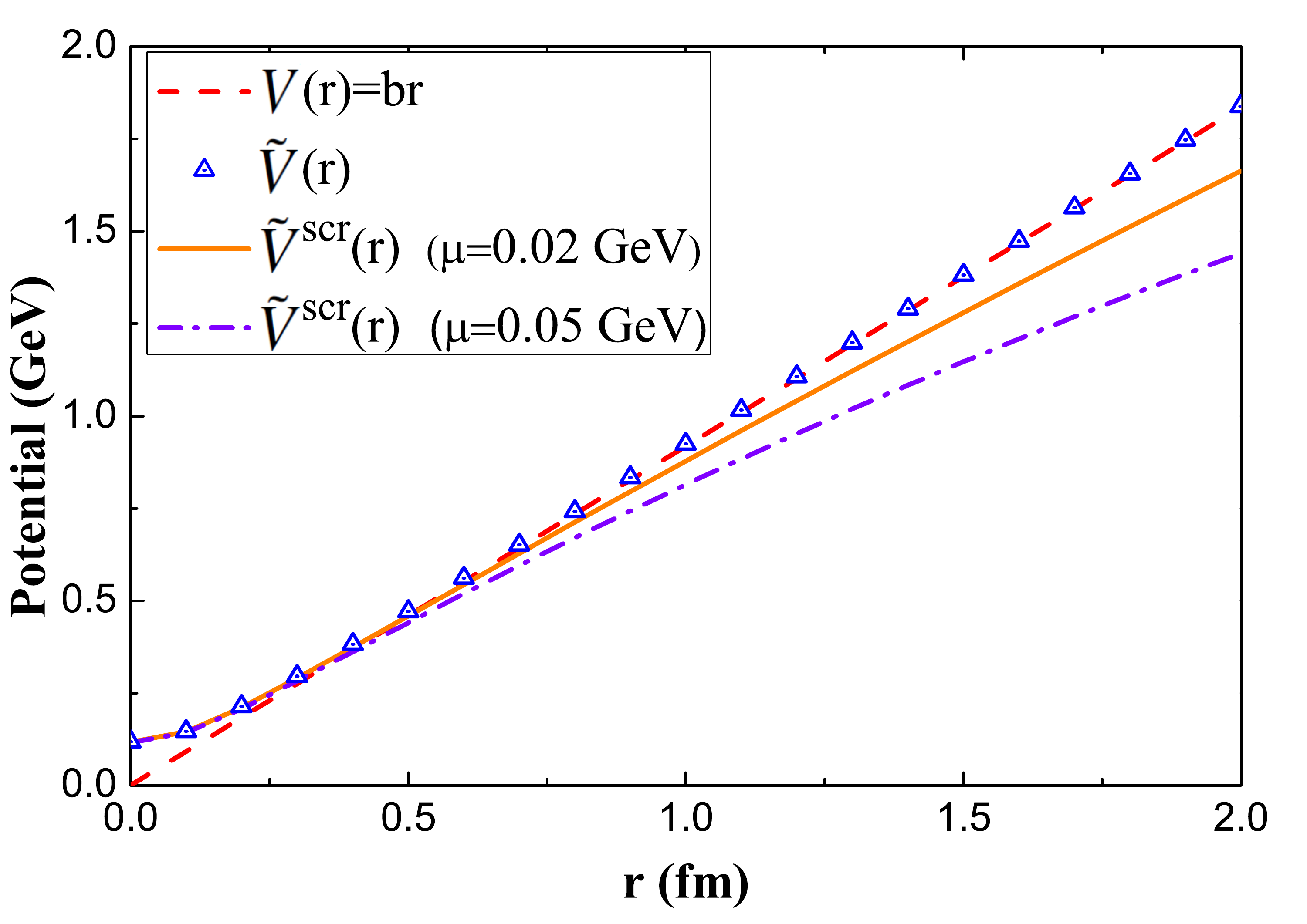}
\caption{\label{fig:muon}(color online). The $r$ dependence of $\tilde V^{\mathrm{scr}}(r)$, $V(r)=br$ and $\tilde V(r)= \int d^3 \bm{r}^\prime
\rho_{12}(\bm{r-r^\prime})br'$. Here, we take two values $\mu=0.02$ GeV and $\mu=0.05$ GeV to show $\tilde V^{\mathrm{scr}}(r)$. Other parameters involved in $\tilde V^{\mathrm{scr}}(r)$, $V(r)=br$ and $\tilde V(r)$ are $m_s=1.628$ GeV, $m_c=0.419$ GeV, $\sigma_0=1.80$ GeV and $s=1.55$ in the expression of $\sigma_{12}$. See TABLE II in \cite{Godfrey:1985xj}. }
\end{figure}

\renewcommand{\arraystretch}{1.5}
\begin{table*}[htbp]
\caption{Mass spectrum of the charmed-strange meson family obtained by our modified GI model and the comparison with those calculated by the GI model. Here, we take $\mu=0.01, 0.02, 0.03, \mathrm{and}~ 0.04  ~\mathrm{GeV}$ to show the results with the modified GI model. The values in brackets are the obtained $R=1/\beta$ and $n$ denotes the radial quantum number. We emphasize that we do not consider the mixing among states with the same quantum number when presenting the results in this Table. $\mu$ is in units of GeV, while $R$ is units of GeV$^{-1}$. }
\centering
\begin{tabular}{l c c| c c c c c c c c  }\toprule[1pt]

&\multicolumn{2}{c|}{GI model \cite{Godfrey:1985xj,Godfrey:2013aaa}}&\multicolumn{8}{c}{Modified GI model}\\
&$n=1$&$n=2$&\multicolumn{4}{c}{$n=1$}& \multicolumn{4}{c}{$n=2$} \\
&&&$\mu=0.01$&$\mu=0.02$&$\mu=0.03$& $\mu=0.04$& $\mu=0.01$&$\mu=0.02$&$\mu=0.03$& $\mu=0.04$\\ \midrule[1pt]
${n}^1S_0$&1979(1.41)&2673(2.00)&1971&1967(1.41)&1963&1960&2661&2646(2.08)&2632&2618  \\
${n}^3S_1$&2129(1.69)&2732(2.08)&2120&2115(1.71)&2111&2106&2719&2704(2.17)&2688&2673  \\
${n}^1P_1$&2550(1.92)&3024(2.17)&2541&2531(1.96)&2522&2512&3001&2979(2.27)&2957&2935  \\
${n}^3P_0$&2484(1.75)&3005(2.13)&2473&2463(1.79)&2454&2444&2982&2960(2.22)&2937&2914  \\
${n}^3P_1$&2552(1.89)&3033(2.22)&2542&2532(1.96)&2522&2512&3010&2988(2.27)&2965&2942  \\
${n}^3P_2$&2592(2.08)&3048(2.27)&2581&2571(2.17)&2561&2551&3026&3004(2.38)&2981&2959  \\
${n}^1D_2$&2910(2.22)&3307(2.38)&2893&2877(2.27)&2861&2844&3277&3247(2.50)&3216&3186  \\
${n}^3D_1$&2899(2.08)&3306(2.27)&2882&2865(2.13)&2848&2831&3275&3244(2.44)&3213&3182  \\
${n}^3D_2$&2916(2.17)&3313(2.38)&2899&2882(2.27)&2865&2848&3283&3252(2.50)&3221&3190  \\
${n}^3D_3$&2917(2.33)&3311(2.44)&2900&2883(2.38)&2867&2850&3281&3251(2.56)&3221&3190  \\
${n}^1F_3$&3199(2.38)&--        &3175&3151(2.50)&3127&3102&--  &--        &--  &--  \\
${n}^3F_2$&3208(2.27)&--        &3183&3159(2.38)&3134&3109&--  &--        &--  &--  \\
${n}^3F_3$&3205(2.33)&--        &3181&3157(2.44)&3132&3107&--  &--        &--  &--  \\
${n}^3F_4$&3190(2.44)&--        &3167&3143(2.56)&3120&3096&--  &--        &--  &--  \\\bottomrule[1pt]
\end {tabular}\\
\label{table:spetrum} 
\end{table*}

\renewcommand{\arraystretch}{1.6}
\begin{table*}[htbp]
\caption{\label{table:comparision}Comparison of the experimental data and theoretical results. Here, we also list the $\chi^2$ values for different models. The notation $L_L$ is introduced to express mixing states either of $^1L_L$ or $^3L_L$.}
\centering
\begin{tabular}{ c c c c c c }\toprule[1pt]
&$n \ ^{2S+1}L_J$ &Experimental values \cite{Beringer:1900zz} & GI model \cite{Godfrey:1985xj,Godfrey:2013aaa}&Modified GI model\footnote{The results listed in the last column are calculated via the modified GI model with $\mu=0.02 $ GeV.} \\ \midrule[1pt]
$D_s$ &$1 \ ^1S_0$&$1968.49\pm0.33$&1979&1967 \\
$D_s^{\ast}$&$1 \ ^3S_1$&$2112.3\pm0.5$&2129& 2115 \\
$D_{s1}(2536)$&$1 \ P_1$&$2535.12\pm0.13$&2556&2534 \\
$D_{s2}^\ast(2573)$&$1 \ ^3P_2$&$2571.9\pm0.8$&2592& 2571\\
$D_{s1}^\ast(2700)$&$2 \ ^3S_1$&$2709\pm4$&2732& 2704\\
$D_{s1}^\ast(2860)$&$1 \ ^3D_1$&$2859\pm12\pm6\pm23$ \cite{Aaij:2014xza,Aaij:2014baa}&2899& 2865\\
$D_{s3}^\ast(2860)$&$1 \ ^3D_3$&$2860.5\pm 2.6 \pm 2.5 \pm 6.0 $ \cite{Aaij:2014xza,Aaij:2014baa}&2917&2883 \\
$D_{sJ}(3040)$&$2 \ P_1$&$3044\pm8^{+30}_{-5}$ \cite{Aubert:2009ah}&3038& 2992\\ \midrule[1pt]
$\chi^2$&--&--&7165&36 \\ \bottomrule[1pt]
\end {tabular}
\label{table:compare}
\end{table*}

\begin{figure*}[htpb]
\includegraphics[width=0.9\textwidth]{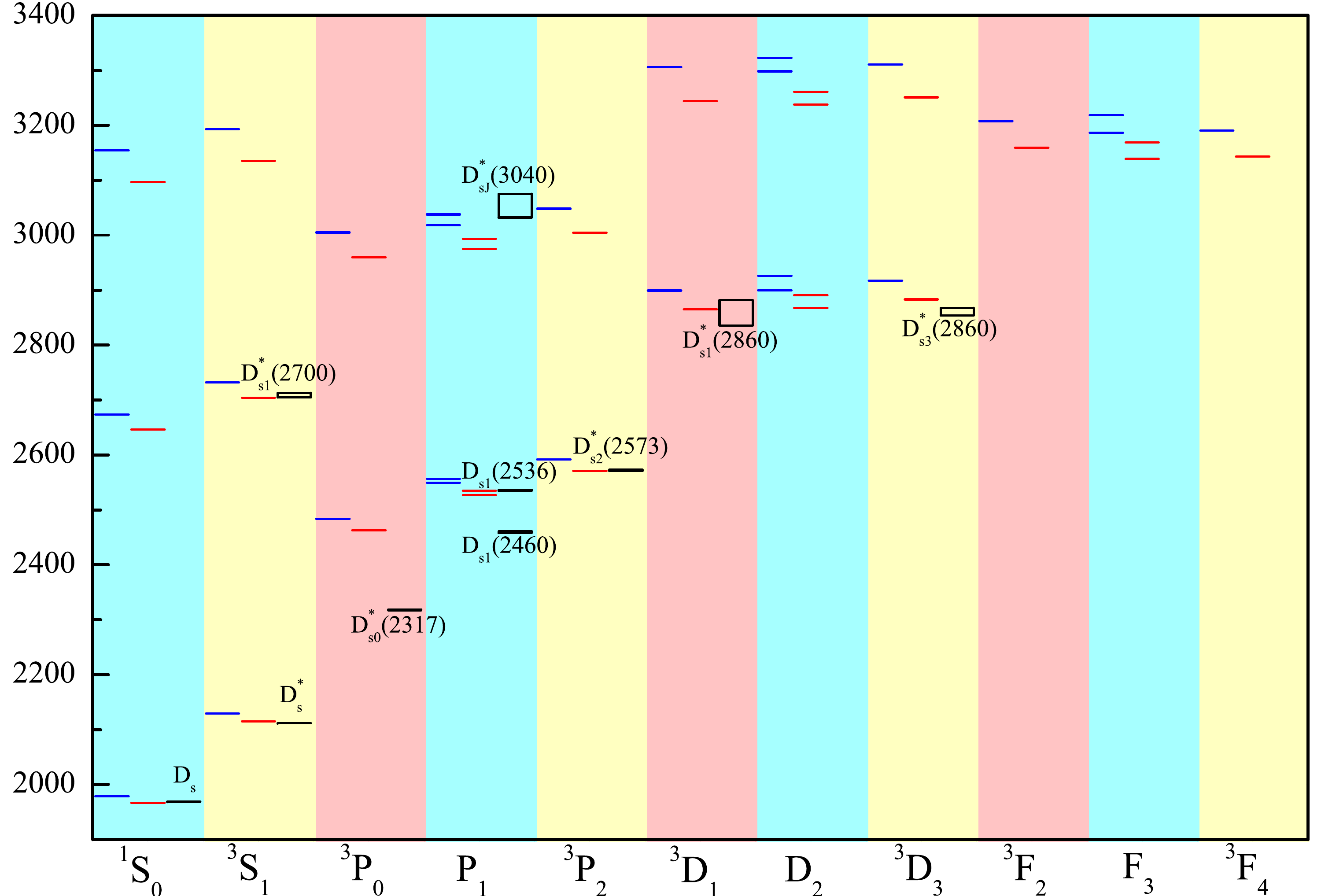}
\caption{\label{fig:spectrum} (color online). Mass spectrum of charmed-strange mesons (in units of MeV).
Here, the blue lines stand for the results obtained by the GI model \cite{Godfrey:1985xj,Godfrey:2013aaa}, while the red lines are those by the modified GI model with $\mu=0.02$ GeV. The black rectangles denote the experimental data taken from PDG \cite{Beringer:1900zz}. The corresponding $^{2S+1}L_J$ quantum numbers are listed in the abscissa. In addition, when there exists
a mixture of $n^1L_L$ and $n^3L_L$ states, we use another notation $L_L$.}
\end{figure*}

In Fig. \ref{fig:muon}, we compare the $r$ dependence of our $\tilde V^{\mathrm{scr}}(r)$, usual linear potential $V(r)=br$ and $\tilde V(r)= \int d^3 \bm{r}^\prime
\rho_{12}(\bm{r-r^\prime})br^\prime $\footnote{The concrete expression for $\tilde V(r)$ is given in Ref. \cite{Godfrey:1985xj} as $$
\tilde V(r)=br\left[\frac{e^{-\sigma^2r^2}}{\sqrt{\pi} \sigma r }+\left(1+\frac{1}{2\sigma^2r^2}\right)\frac{2}{\sqrt{\pi}}\int_0^{\sigma r}e^{-x^2}dx\right].$$}. While behaviors of $\tilde V^{\mathrm{scr}}(r)$ and $\tilde V(r)$ are similar to each other at small distances, there exists difference between $V(r)=br$ and smeared $\tilde V(r)$ reflecting the relativistic effect. There is obvious difference between $\tilde V^{\mathrm{scr}}(r)$ and $\tilde V(r)$ at large distances due to the screening effect. {Comparing a heavy-light system with a heavy quarkonium, a velocity of a light quark in the heavy light meson is much larger than that of a heavy quark in the quarkonium and in addition, a radius of the heavy light meson is also larger than the heavy quarkonium. These facts indicate that we need to consider the relativistic and screening effects when studying the mass spectra of higher radial and orbital excitations of charmed-strange mesons.} Furthermore, adjusting $\mu$ is to control the power of screening effect.

\subsection{Numerical results}

In the following, we present numerical results of the mass spectrum of the charmed-strange meson family.
In Table \ref{table:spetrum}, the masses of the charmed-strange mesons are listed using the GI model \cite{Godfrey:1985xj} and our modified GI model, where we take several $\mu$ values in $\tilde V^{\mathrm{scr}}(r)$ to show $\mu$ dependency of our model. Apart from $\mu$, other parameters appearing in our calculation are taken from Ref. \cite{Godfrey:1985xj},
with which we can, of course, reproduce the masses obtained in Refs. \cite{Godfrey:1985xj,Godfrey:2013aaa}. In Table \ref{table:spetrum}, we also give the $R=1/\beta$ values corresponding to charmed-strange mesons, where the $R$ value can be obtained by the relation
\begin{eqnarray}
\int\Psi^{\mathrm{SHO}}_{nLM}(\mathbf{p})^2 p^2 d^3\mathbf{p}=\int\Phi(\mathbf{p})^2 p^2 d^3\mathbf{p}.\label{h2}
\end{eqnarray}
The r.h.s of Eq. (\ref{h2}) is the root mean square momentum, which can be directly calculated through the GI model or the modified GI model. The l.h.s of Eq. (\ref{h2}) is a definition of the root mean square momentum when adopting the SHO wave function.
Here, in the momentum space the SHO wave function is expressed as
\begin{eqnarray}
\Psi^{\mathrm{SHO}}_{nLM_L}(\bm{p})=R^{\mathrm{SHO}}_{nL}(p)Y_{LM_L}(\Omega_p),
\end{eqnarray}
with the radial wave function
\begin{eqnarray}
R^{\mathrm{SHO}}_{nL}(p)&=&\frac{(-1)^n(-i)^L}{\beta^{3/2}}\sqrt{\frac{2n!}{\Gamma(n+L+3/2)}}\left(\frac{p}{\beta}\right)^L
e^{-\frac{p^2}{2\beta^2}} \nonumber \\ &&\times L_n^{L+1/2}(p^2/\beta^2),\label{k11}
\end{eqnarray}
where $L_n^{L+1/2}(p^2/\beta^2)$ is an associated Laguerre polynomial with the oscillator parameter $\beta$.
In the configuration space the SHO wave function is given by
\begin{eqnarray}
\Psi^{\mathrm{SHO}}_{nLM_L}(\bm{r})=R^{SHO}_{nL}(r)Y_{LM_L}(\Omega_r),
\end{eqnarray}
where the radial wave function is defined as
\begin{eqnarray}
R^{\mathrm{SHO}}_{nL}(r)=\beta^{3/2}\sqrt{\frac{2n!}{\Gamma(n+L+3/2)}}(\beta r )^L e^{\frac{-\beta^2r^2}{2\beta^2}}L_n^{L+1/2}(\beta^2r^2).
\end{eqnarray}
The $\beta$ is the parameter appearing in the SHO radial wave function given in Eq. (\ref{k11}), which is determined by the above procedure.

In Table \ref{table:comparision}, we further compare the calculated results with the experimental data. Illustrating further a suitable $\mu$ value
{introduced in the modified GI model}, we also list the $\chi^2$ values for the GI model and the modified GI model, where $\chi^2$ is defined as
\begin{eqnarray}
\chi^2=\sum_i\left(\frac{\mathcal{A}_{\mathrm{Th}}(i)-\mathcal{A}_{\mathrm{Exp}}(i)}{\mathrm{Error}(i)}\right)^2,
\end{eqnarray}
where $\mathcal{A}_{\mathrm{Th}}(i)$ and $\mathcal{A}_{\mathrm{Exp}}(i)$ are theoretical and experimental values.
$\mathrm{Error}(i)$ is the experimental error listed in Table \ref{table:comparision}.
In Table \ref{table:comparision}, the values of $\chi^2$ are calculated without the contributions from $D_{s0}(2317)$ and $D_{s1}(2460)$. Comparing the $\chi^2$ values for different cases, we find that the modified GI model can well describe the experiments when $\mu=0.02$ GeV and that the description of the observed charmed-strange spectrum can be obvisously improved by the modified GI model with the screening effect. {One can clearly see the trend of decreasing masses with increasing $\mu$ from Table \ref{table:comparision}.
}

In Fig. \ref{fig:spectrum}, we further list the results obtained by the modified GI model ($\mu=0.02$ GeV) and make a comparison with the GI model and experimental data. As for $D_s$, $D_s^{\ast}$, $D_{s1}(2536)$, $D_{s2}^\ast(2573)$, $D_{s1}^\ast(2700)$ and the newly observed $D_{s1}^\ast(2860)$, the mass differences between the experimental data and our theoretical results are less than 10 MeV.
While the theoretical mass of $D_{sJ}^\ast(2860)$ is about 20 MeV lower than the experimental value,
we notice that the experimental masses of $D_{s0}^\ast(2317)$ and $D_{s1}(2460)$ can not be reproduced by the modified GI model. This discrepancy is caused by the near-threshold effect which is ignored in the screening effect \cite{Li:2009zu}.
In general, the comparison of 10 experimental data with our calculated results indicates that the modified GI model is more suitable to describe the experimental data, especially to higher charmed-strange mesons.

We should emphasize that the $n^1P_1-n^3P_1$ and $n^1D_2-n^3D_2$ mixtures with $n=1,2$ are included in the corresponding calculations listed in Table \ref{table:comparision} and Fig. \ref{fig:spectrum}, which is different form the situation in Table \ref{table:spetrum} since we need to compare theoretical results with the experimental data.

\section{Two-body OZI-allowed strong decays}\label{sec4}

In addition to the mass spectrum, the decay properties are also crucial features of mesons. The quarks pair creation (QPC) model is successful to calculate Okubo-Zweig-Iizuka (OZI) allowed strong decays of mesons.
Here, we firstly give a brief introduction to the QPC model.

\subsection{Brief introduction to the QPC model }

The QPC model was proposed by Micu \cite{Micu:1968mk} and developed by the Orsay group \cite{Le Yaouanc:1972ae,LeYaouanc:1988fx,vanBeveren:1979bd,vanBeveren:1982qb,Bonnaz:2001aj,roberts}. It assumes that meson  decay occurs through a flavor-singlet and color-singlet quark-antiquark pair created from the vacuum. The transition operator $T$  can be expressed as

\begin{eqnarray}
T&=&-3\gamma \sum \limits_{m} \langle1m;1-m|00\rangle \int d\mathbf{p}_3 d\mathbf{p}_4 \delta^3(\mathbf{p}_3+\mathbf{p}_4)\nonumber\\
&\times& \mathcal{Y}_{1m}\left(\frac{\mathbf{p}_3-\mathbf{p}_4}{2}\right) \chi^{34}_{1,-m} \phi^{34}_{0} \left(\omega^{34}_{0}\right)_{ij}b^{\dag}_{3i}(\mathbf{p}_3) d^{\dag}_{4j}(\mathbf{p}_4),
\end{eqnarray}
where $\mathbf{p}_3$  and  $\mathbf{p}_4$ denote the momenta of quark and antiquark created from the vacuum, respectively. $\gamma$ is the parameter  which describes the strength of the creation of quark-antiquark pair. By comparing the experimental widths with theoretical ones of 16 famous decay channels, $\gamma=8.7$ is obtained for the $u\bar{u}/d\bar{d}$ pair creation \cite{Ye:2012gu}.
For the $s\bar{s}$ pair creation, we take $\gamma=8.7/\sqrt{3}$ \cite{Le Yaouanc:1972ae}.
$\phi^{34}_{0}=(u \bar{u}+d \bar{d}+s  \bar{s})/\sqrt{3}$ is the flavor function while  $\left(\omega^{34}_{0}\right)_{ij}=\delta_{ij}/\sqrt{3}$ is the color function, where $i$ and $j$ are the color indices. $\mathcal{Y}_{\ell m}(\mathbf{p})=|\mathbf{p}|^{\ell} Y_{\ell m}(\mathbf{p})$ denotes the solid harmonic polynomial. Thus,  $\mathcal{Y}_{1m}(\frac{\mathbf{p}_3-\mathbf{p}_4}{2})$ indicates that the angular momentum of the quark-antiquark pair is $L=1$. The  $\chi^{34}_{1,-m}$ means that the total spin angular momentum of the quark-antiquark pair is $S=1$. Finally, the quantum number of the quark-antiquark pair is contained as $J^{PC}=0^{++}$ through coupling of the angular momentum with the spin angular momentum.

The transition matrix of meson $A$ decaying into mesons $B$ and $C$ in the $A$ rest frame is defined as
\begin{eqnarray}
\langle BC|T|A\rangle=\delta^3(\mathbf{p}_B+\mathbf{p}_C)\mathcal{M}^{M_{JA}{M_{JB}}{M_{JC}}},
\end{eqnarray}
$\mathbf{p}_B$ and $\mathbf{p}_C$ are the momenta of mesons $B$ and $C$, respectively. $|A\rangle$, $|B\rangle$ and $|C\rangle$ denote the mock states \cite{Hayne:1981zy}. Taking a meson A as an example, we illustrate the definition of a mock state, i.e.,
\begin{eqnarray}
&&|A(n^{2S+1}L_{JM_J})(\mathbf{p}_A)\rangle\nonumber\\
&&=\sqrt{2E}\sum\limits_{ {M_{S}},{M_{L}}}\langle LM_L;SM_{S}|JM_{J}\rangle\chi^{A}_{S,M_s} \nonumber\\
&&\quad\times \phi^A \omega^A\int d\mathbf{p}_1 d\mathbf{p}_2 \delta^3(
\mathbf{p}_A-\mathbf{p}_1-\mathbf{p}_2) \nonumber\\ &&\quad\times\Psi^A_{nLM_L}(\mathbf{p}_1,\mathbf{p}_2)|q_1(\mathbf{p}_1)\bar{q}_2(\mathbf{p}_2)\rangle,
\end{eqnarray}
where $\chi^{A}_{S,M_s}$, $\phi^A$ and $\omega^A$ are the spin, flavor and color wave functions, respectively. $\Psi^A_{nLM_L}(\mathbf{p}_1,\mathbf{p}_2)$ is the spacial wave function of a meson A in the momentum space, which can be obtained by the modified GI model. The calculated amplitude $M^{M_{JA}{M_{JB}}{M_{JC}}}$ can be converted into the partial wave amplitude $M^{JL}$ via the Jacob-Wick formula \cite{Jacob:1959at}, i.e.,
\begin{eqnarray}
\mathcal{M}^{JL}(A\rightarrow BC)&=&\frac{\sqrt{2L+1}}{2J_A+1}\sum \limits_{ {M_{JB}},{M_{JC}}}\langle L0;JM_{JA}|J_AM_{JA}\rangle\nonumber\\&&\times
\langle J_BM_{JB};J_CM_{JC}|JM_{JA}\rangle \mathcal{M}^{M_{JA}{M_{JB}}{M_{JC}}}.\nonumber\\
\end{eqnarray}
Finally, the decay width can be expressed as
\begin{eqnarray}
\Gamma=\pi^2\frac{|\mathbf{p}_B|}{m_A^2}\sum \limits_{J,L}|\mathcal{M}^{JL}|^2.
\end{eqnarray}

We would like to emphasize the improvement of the present work compared with the former works in Refs. \cite{Sun:2009tg,Zhang:2006yj,Song:2014mha}. In Refs. \cite{Sun:2009tg,Zhang:2006yj,Song:2014mha}, authors adopted the SHO wave function to describe a spacial wave function of charmed and charmed-strange mesons, where the $\beta$ value in the SHO wave function is determined by Eq. (\ref{h2}). This treatment is an approximation for simplifying the calculation since the SHO wave function $\Psi^{\mathrm{SHO}}_{nLM_L}(\bm{p})$ is slightly different from the wave function $\Phi(\bm{p})$ in Eq. (\ref{h2}). In this work, for the charmed-strange meson $A$ involved in our calculation, we use the numerical wave function directly obtained as an eigenfunction of the modified GI model, which can avoid the uncertainty from the treatment in Eq. (\ref{h2}). For the mesons $B$ and $C$ in the discussed process, we still adopt
the SHO wave function, where the corresponding $\beta$ was calculated in Ref. \cite{Godfrey:1986wj}.

In the following concrete calculation, the mass is taken from PDG \cite{Beringer:1900zz} if there exists the corresponding experimental observation of a charmed-strange meson. If the discussed charmed-strange meson is still missing, we adopt the theoretical prediction from the modified GI model (the results listed in Fig. \ref{fig:spectrum}) as the input.

\subsection{Phenomenological analysis of strong decay}

In this subsection, we carry out the phenomenological analysis of the strong decays of charmed-strange mesons by combining our theoretical results with the corresponding experimental data.

\renewcommand{\arraystretch}{1.6}
\begin{table*}[htbp]
\caption{Partial widths of $D_{s0}^*(2317)\to D_s \pi^0$ and $D_{s1}(2460)\to D_s^*\pi^0$ (in the unit of keV). Here, we list the width of $D_{s1}(2460)\to D_s^*\pi^0$ with a mixing angle $\theta_{1P}=-54.7^\circ=-\arcsin(\sqrt{2/3})$ in the heavy quark limit. }
\centering
\begin{tabular}{ c c c c c c c c c c c c}\toprule[1pt]
  &this work &Ref. \cite{Lu:2006ry} &Ref.\cite{Bardeen:2003kt} &  Ref. \cite{Fayyazuddin:2003aa}&Ref. \cite{Godfrey:2003kg}&Ref. \cite{Wei:2005ag}& Ref. \cite{Colangelo:2003vg}&Ref. \cite{Cheng:2003kg}&Ref. \cite{Ishida:2003gu} &Ref. \cite{Matsuki:2011xp}\\ \midrule[1pt]
$\Gamma(D_{s0}^\ast(2317)\rightarrow D_s \pi)$ &11.7 &32&21.5&16&$\sim$10&34-44&$\simeq6$&10-100&$155\pm70$&3.8\\
$\Gamma(D_{s1}(2460)\rightarrow D^\ast_s \pi)$ &11.9&35&21.5&32&$\sim$10&35-51&$\simeq6$&--&$155\pm70$&3.9\\ \bottomrule[1pt]
\end {tabular}
\label{table:singlepion}
\end{table*}

\subsubsection{$1P$ states}

\begin{figure}[htpb]
\includegraphics[width=0.45\textwidth]{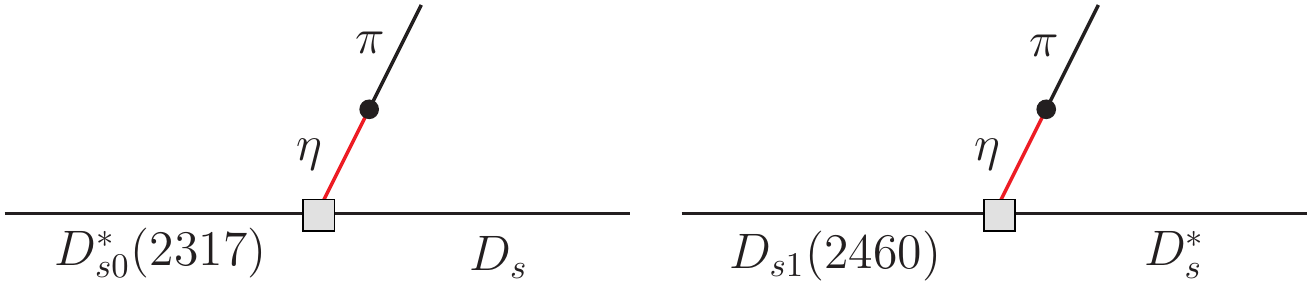}
\caption{\label{fig:etapi} The $D_{s0}^*(2317)\to D_s \pi^0$ (left) and $D_{s1}(2460)\to D_s^*\pi^0$ (right) decays through the $\eta-\pi^0$ mixing.}
\end{figure}

If $D_{s0}^\ast(2317)$ and $ D_{s1}(2460)$ are the $1P$ states in the charmed-strange meson family, the OZI-allowed strong decays are forbidden since their masses are below the $DK/D^*K$ thresholds. However, $D_{s0}^*(2317)\to D_s \pi^0$ and $D_{s1}(2460)\to D_s^*\pi^0$ can occur via the $\eta-\pi$ mixing, which are shown in Fig. \ref{fig:etapi} \cite{Lu:2006ry}.

For calculating these decays, we need the Lagrangian including the $\eta-\pi^0$ mixing, which can be expressed as \cite{Cho:1994zu}
\begin{eqnarray}
\mathcal{L}_{\eta-\pi^0}=\frac{m_{\pi}^2f^2}{4(m_u+m_d)}\mathrm{Tr}(\xi m_q \xi+\xi^\dag m_q \xi^\dag),
\end{eqnarray}
where $m_q$ is a light quark mass matrix,
 $\xi=\mathrm{exp}(i\tilde{\pi}/\mathrm{f}_{\pi})$ and $\tilde{\pi}$
is a light meson octet.

Taking $D_{s0}^*(2317)\rightarrow D_s\pi^{0}$ as an example, we illustrate the concrete calculation. The decay amplitude of
$D_{s0}^*(2317)\rightarrow D_s\pi^{0}$ reads as
\begin{eqnarray}
\mathcal{M}^{JL}(D_{s0}^*(2317)\to D_s\pi^{0})&=&\mathcal{M}^{JL}(D_{s0}^*(2317)\to D_s\eta)\nonumber\\
&&\times\frac{i\sqrt{3}}{4}\delta_{\pi^0 \eta},\label{pion}
\end{eqnarray}
where decay amplitude $\mathcal{M}^{JL}(D_{s0}^*(2317)\to D_s\eta)$ can be calculated
by the QPC model. The isospin-violating factor $\delta_{\pi^0
\eta}$ is due to the $\eta-\pi$ mixing, i.e., \cite{Gasser:1984gg}
\begin{eqnarray}
\delta_{\pi^0 \eta}=\frac{m_d-m_u}{m_s-(m_u+m_d)/2}=\frac{1}{43.7}.\label{iso}
\end{eqnarray}
Adopting the similar treatment, we can derive the formula of $D_{s1}(2460)\to D_s^*\pi^0$.
Here, $D_{s1}(2460)$ with $J^P=1^+$ is the mixture between $1^1P_1$ and $1^3P_1$ states, which satisfies the relation
\begin{equation}
 \left(
  \begin{array}{c}
   |D_{s1}(2460)\rangle\\
   |D_{s1}(2536)\rangle\\
  \end{array}
\right )=
\left(
  \begin{array}{cc}
    \cos\theta_{1P} & \sin\theta_{1P} \\
   -\sin\theta_{1P} & \cos\theta_{1P}\\
  \end{array}
\right)
\left(
  \begin{array}{c}
    |1^1P_1  \rangle \\
   |1^3P_1 \rangle\\
  \end{array}
\right),\label{m1}
\end{equation}
where the mixing angle $\theta_{1P}=-54.7^\circ$ is obtained in the heavy quark limit \cite{Godfrey:1986wj,Matsuki:2010zy,Barnes:2005pb}.

The obtained partial widths of $D_{s0}^*(2317)\to D_s \pi^0$ and $D_{s1}(2460)\to D_s^*\pi^0$ are listed in Table \ref{table:singlepion}. We also compare our results with other theoretical predictions from different groups \cite{Lu:2006ry,Bardeen:2003kt,Fayyazuddin:2003aa,Godfrey:2003kg,Wei:2005ag,Colangelo:2003vg, Cheng:2003kg,Ishida:2003gu,Matsuki:2011xp}. Especially, we notice that our results are consistent with the corresponding values given in Ref.  \cite{Godfrey:2003kg}.
The equality of these two decay widths predicted in Ref. \cite{Matsuki:2011xp} in the heavy quark limit is well satisfied in our case, too.

\begin{figure}[htbp]
\includegraphics[width=0.44\textwidth]{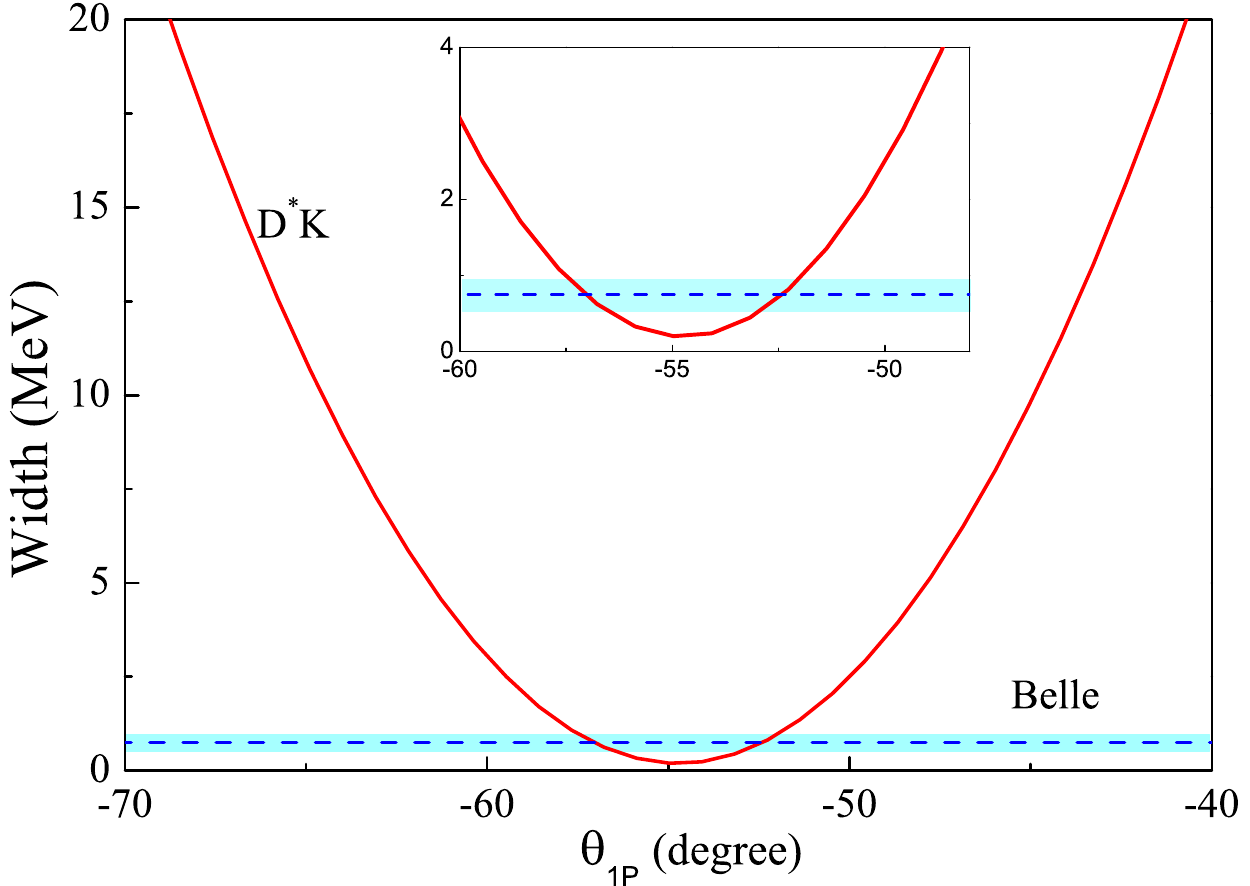}
\caption{\label{fig:ds2536}The mixing angle $\theta_{1P}$ dependence of a decay width $D_{s1}(2536)\to D^{*}K$.}
\end{figure}

$D_{s1}(2536)$ with $J^P=1^+$ is the orthogonal partner of $D_{s1}(2460)$ as shown in Eq. (\ref{m1}). The $D^*K$ channel is its only OZI allowed decay mode. In Fig. \ref{fig:ds2536}, we present the dependence of the partial decay width of
$D_{s1}(2536)\to D^*K$ on $\theta_{1P}$, which covers the typical value $\theta_{1P}=-54.7^\circ$ given in the heavy quark limit \cite{Godfrey:1986wj,Matsuki:2010zy,Barnes:2005pb}. We notice that our result is consistent with the experimental data $\Gamma=0.92\pm0.05$ MeV \cite{Lees:2011um} and $\Gamma=0.75\pm0.23$ MeV \cite{Belle:2011ad}, when the mixing angle is around $\theta_{1p}=-52.3^\circ$ and $\theta_{1p}=-53.1^\circ$, respectively, which are close to $\theta_{1p}=-54.7^\circ$ of the heavy quark limit. 
{In the heavy quark limit, the decay of $D_{s1}(2536)$ into $D^\ast K$ can occur only via a $D$ wave because of the conservation of light degree of freedom and hence the decay width is expected to be small as shown in Fig. \ref{fig:ds2536}. That is, because the experimental mixing angle is very close to the one of the heavy quark limit, it is a model independent result that $D_{s1}(2536)$ and $D^\ast K$ almost exactly decouples with each other, which has been discussed in Refs. \cite{Godfrey:1986wj, Rosner:1985dx} } 

\renewcommand{\arraystretch}{1.3}
\begin{table*}[htb]
\caption{\label{table:channel}Calculated partial decay widths of the OZI-allowed strong decays of $2S$, $1D$, $3S$, and $2P$ charmed-strange mesons. Units are in MeV. Here, forbidden decay channels are
marked by -- and channels marked by $\Box$ are the OZI-allowed modes, which are discussed in consideration of the mixing angle dependence.}
\centering
\begin{tabular}{c c c c c c c c c c c c c c}\toprule[1pt]

\multirow{2}{*}{Channels}&
\multirow{2}{*}{$D_s(2^1S_0)$}&\multirow{2}{*}{$D^*_{s1}(2700)$}&\multirow{2}{*}{$D^*_{s1}(2860)$}&$D_s(1D(2^-))$&\multirow{2}{*}{$D_{s3}^*(2860)$}&
\multirow{2}{*}{$D_s(2^3P_0)$}    & \multirow{2}{*}{$D_s(2P(1^+))$}&\multirow{2}{*}{$D_s(2P'(1^+))$} &\multirow{2}{*}{$D_s(2^3P_2)$}& \multirow{2}{*}{$D_s(3^1S_{0})$} & \multirow{2}{*}{$D_s(3^3S_1)$}\\
                     &     &      &      &$D_s(1D'(2^-))$&     &     &   &          &                     &     &\\ \midrule[1pt]
$D K$                  &--   &$\Box$&$\Box$&--        &7.46 &60.86&--& --        &$2.99\times 10^{-4}$ &--   &23.30 \\
$D^{*}K$             &76.06&$\Box$&$\Box$&$\Box$    &5.98 &--   &62.43&20.72   &5.30                 & 38.11  &36.54\\
$D_s\eta$            &--   &$\Box$&$\Box$&--        &0.15 &2.07 &--&--         &0.02                 & --  &2.19\\
$D_{s}\eta'$         &--   &--    &--    &--        &--   &0.08 &--& --        &0.001                &  -- &0.40\\
$D_{s}^*\eta$        &--   &$\Box$&$\Box$&$\Box$    &0.06 &--   &1.61& 1.02    &0.18                 &2.56&2.72\\
$D_{s}^*\eta^\prime$ &--   &--    &--    &--        &--   &--   &--&--         &--                   &  --  &0.07\\
$DK^*$               &--   &--    &$\Box$&$\Box$    &0.37 &--   &39.32&6.50    &13.46                &3.33 &14.92\\
$D^*K^*$             &--   &--    &--    &--        &--   &63.66&48.97&39.60   &48.32                &18.54  &$3.30\times 10^{-5}$\\
$D_s\phi$            &--   &--    &--    &--        &--   &--   &1.45&8.05     &0.003                &0.03   &0.07\\
$D_s^*\phi$          &--   &--    &--    &--        &--   &--   &--&--         &--                   & --     &$9.97\times 10^{-4}$\\
$D_0^*(2400)K$       &--   &--    &--    &$\Box$    &--   &--   &14.42&31.13   &--                   & 38.14   &--\\
$D_{s0}^*(2317)\eta$ &--   &--    &--    &$\Box$    &--   &--   &1.06& 2.41    &--                   &5.38        &--\\
$D_1(2430)K$         &--   &--    &--    &--        &--   &3.32 &14.31&8.11    &14.31                &  --      &17.02\\
$D_1(2420)K$         &--   &--    &--    &--        &--   &36.16&19.43&10.43   &0.75                 &  --      &5.50\\
$D_{s1}(2460)\eta$   &--   &--    &--    &--        &--   &--   &0.25&0.16     &--                   &  --      &3.43\\
$D_{s1}(2536)\eta$   &--   &--    &--    &--        &--   &--   &--&--         &--                   &   --     &0.02\\
$D_2^*(2460)K$       &--   &--    &--    &--        &--   &--   &82.85&3.15    &3.90                 &5.89    &9.16\\
$D_{s2}^*(2573)\eta$ &--   &--    &--    &--        &--   &--   &--&--         &--                   &  --      &0.002\\ \midrule[1pt]
Total                &76.06&--    &--    &--        &14.02&166.15&285.83&131.28&86.25                &111.98    &115.35 \\ \bottomrule[1pt]
\end {tabular}\\
\label{table:decay}
\end{table*}

As a good candidate of the charmed-strange meson with $1^3P_2$, $D_{s2}(2573)$ decays into $DK$, $D^*K$, and $D_s\eta$ with the partial decay widths $5.42$ MeV, $0.57$ MeV, and $0.04$ MeV, respectively. The total decay width of $D_{s2}(2573)$ can reach up to 6.03 MeV, which is comparable with the experimental value $16^{+5}_{-4}\pm3$ \cite{Kubota:1994gn}, $10.4\pm8.3\pm3.0$ MeV \cite{Albrecht:1995qx} and $12.1\pm4.5\pm1.6$ MeV \cite{Aaij:2011ju} given by the ARGUS collaboration and LHCb collaboration, respectively. In addition, the branching ratio
\begin{eqnarray}
\frac{\mathcal{B}(D_{s2}(2573)\rightarrow
D^{\ast0}K^+)}{\mathcal{B}(D_{s2}(2573)\rightarrow D^0K^+)}<0.33
\end{eqnarray}
is also measured by the CLEO collaboration \cite{Kubota:1994gn}. In this work, we obtain $\mathcal{B}(D^{\ast0}K^+)/\mathcal{B}(D^0K^+)=0.106$ consistent with the present experimental measurement \cite{Kubota:1994gn}. This value is also close to the one, 0.076, predicted in Ref.~\cite{Matsuki:2011xp}.

\subsubsection{$2S$ and $1D$ states}

Since there is no candidate for $2^1S_0$ charmed-strange meson observed in experiment,
we adopt the theoretical value calculatded by the modified GI model as an input, i.e., 2646 MeV. As shown in Table \ref{table:channel}, $D_s(2^1S_0)$ only decays into $D^*K$, where the decay width of  $D_s(2^1S_0)\to D^*K$ is 76.06 MeV.

In the following, we shall discuss two observed $D_{s1}^*(2700)$ and $D_{s1}^{*}(2860)$ to be an admixture of $D_s(2^3S_1)$ and $D_{s}(1^3D_1)$, which satisfies the relation
\begin{equation}
 \left(
  \begin{array}{c}
   |D_{s1}^*(2700)\rangle\\
   |D_{s1}^{*}(2860)\rangle\\
  \end{array}
\right )=
\left(
  \begin{array}{cc}
    \cos\theta_{SD} & \sin\theta_{SD} \\
   -\sin\theta_{SD} & \cos\theta_{SD}\\
  \end{array}
\right)
\left(
  \begin{array}{c}
    |2^3S_1  \rangle \\
   |1^3D_1 \rangle\\
  \end{array}
\right),\label{sd}
\end{equation}
where $\theta_{SD}$ denotes a mixing angle describing an admixture of $D_s(2^3S_1)$ and $D_{s}(1^3D_1)$.

\begin{figure}[htbp]
\includegraphics[width=0.49\textwidth]{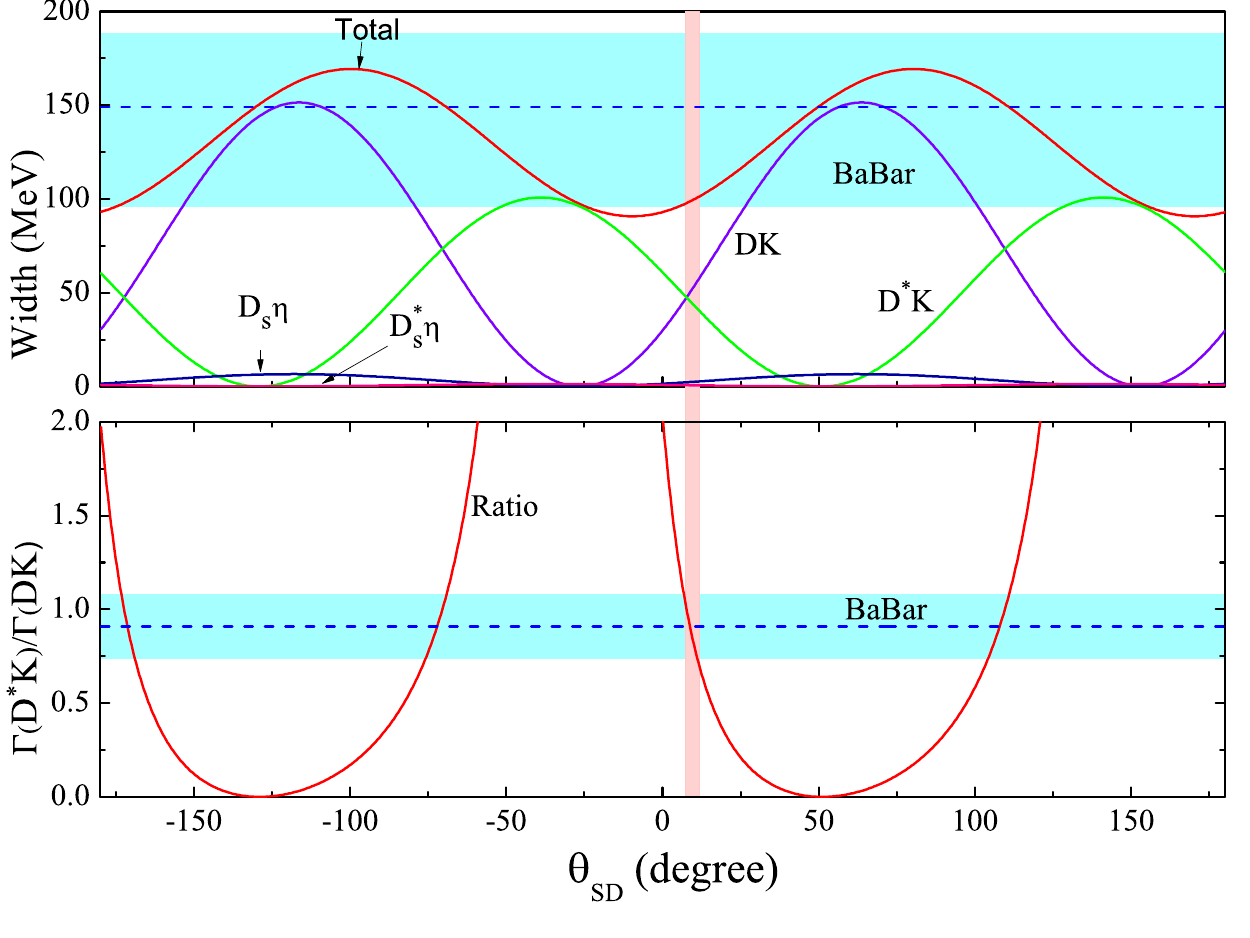}
\caption{\label{fig:ds2700}The $\theta_{SD}$ dependence of the calculated partial and total decay widths and ratio $\Gamma(D^*K)/\Gamma(DK)$. The vertical band corresponds to the common range of $\theta_{SD}$, where our results can be matched with the experimental widths and ratio \cite{Aubert:2009ah}.}
\end{figure}

The $\theta_{SD}$ dependence of the total and partial decay widths of $D_{s1}^\ast(2700)$ is shown in the Fig. \ref{fig:ds2700}. Its main decay modes are $DK$ and $D^*K$, both of which were observed in experiments.
To compare our results with the experimental data, we take the BaBar's measurement in Ref. \cite{Aubert:2009ah} since the width $\Gamma=(149\pm7^{+39}_{-52})$ MeV together with the ratio listed in Eq. (\ref{r1}) were given  \cite{Aubert:2009ah}. As shown in Fig. \ref{fig:ds2700}, there exists the common $\theta_{SD}$ range, $6.8^\circ-11.2^\circ$, in which both of the calculated width and ratio of $D_{s1}^\ast(2700)$ \cite{Aubert:2009ah} can overlap with the experimental data. This small $\theta_{SD}$ value is consistent with the estimate of $\theta_{SD}$ in Ref. \cite{Godfrey:1985xj}.

\begin{figure}[htbp]
\includegraphics[width=0.49\textwidth]{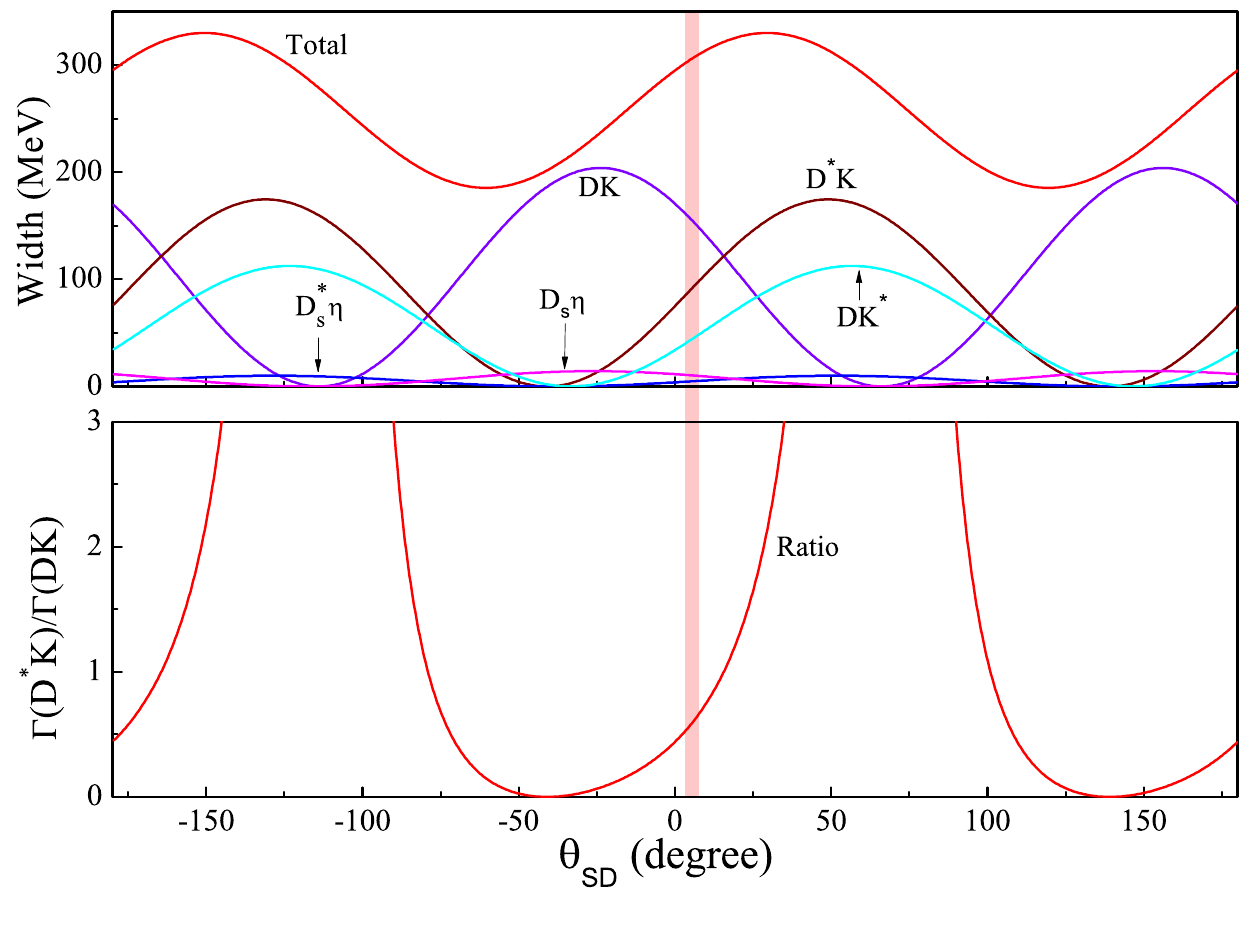}
\caption{\label{fig:ds2860} The $\theta_{SD}$ dependence of the decay widths and ratio $\Gamma(D^*K)/\Gamma(DK)$ of $D_{s1}^*(2860)$. The vertical band corresponds to the common range of $\theta_{SD}$, where our results can be matched with the experimental width and ratio of $D_{s1}^\ast(2700)$ \cite{Aubert:2009ah}. \label{Fig:Ds12860}}
\end{figure}

As for the $D_{s1}^*(2860)$ with $J^P=1^-$ recently observed by the LHCb collaboration \cite{Aaij:2014xza,Aaij:2014baa} which is considered to be a partner of $D_{s1}^\ast(2700)$, the  $\theta_{SD}$ dependence of the decay behavior is depicted in Fig.  \ref{Fig:Ds12860}. If taking $6.8^\circ-11.2^\circ$ for the range of $\theta_{SD}$ obtained in the study of
$D_{s1}^\ast(2700)$, we find that the obtained total decay width of $D_{s1}^*(2860)$ can reach up to $\sim 300$ MeV which is comparable with the LHCb data \cite{Aaij:2014xza,Aaij:2014baa} and the ratio is $\mathcal{B}(D_{s1}^*(2860)\rightarrow D^*K) / \mathcal{B}(D_{s1}^*(2860)\rightarrow DK)=0.6\sim0.8$, which can be tested in future experiment. Our study also shows that the main decay channels of $D_{s1}^*(2860)$ are $DK$($\sim$140 MeV), $D^*K$($\sim$95 MeV) and $DK^*$($\sim$50 MeV) .

\begin{figure}
\includegraphics[width=0.46\textwidth]{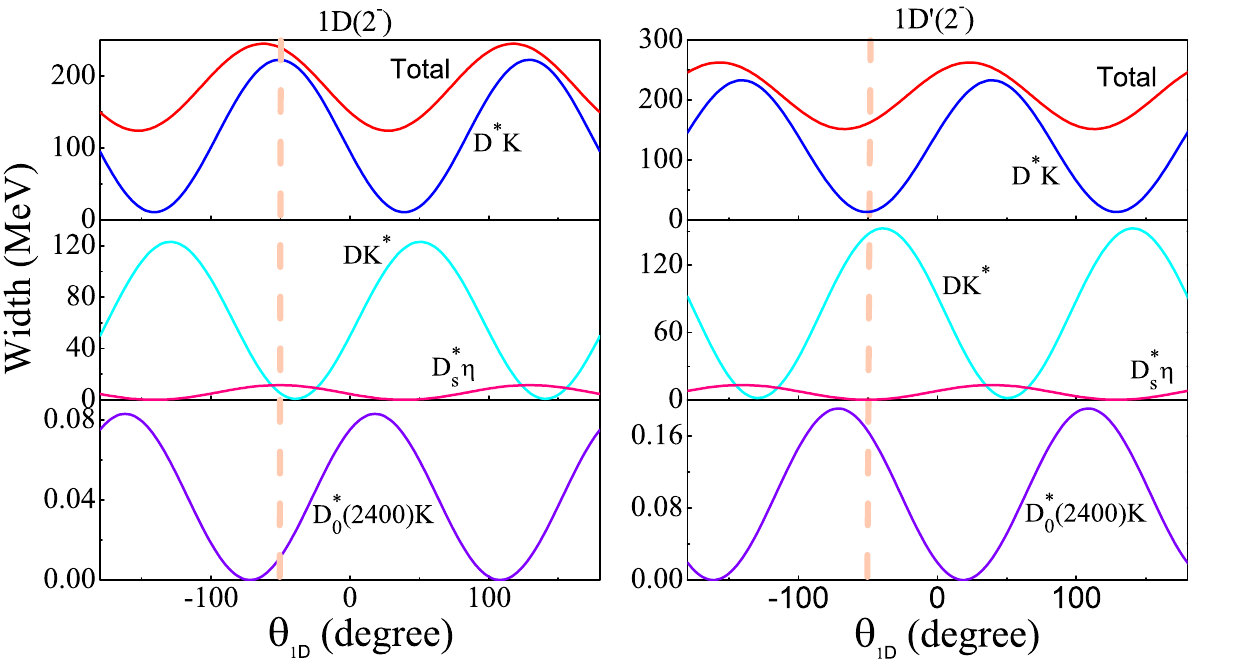}
\caption{\label{fig:1d2} The $\theta_{1D}$ dependence of the partial decay widths and total decay width  of $D_s(1D(2^-))$ (left) and $D_s(1D'(2^-))$ (right). The vertical dash lines
correspond to the mixing angle $\theta_{1D}=-50.8^\circ $}
\end{figure}

As an admixture of  $1^1D_2$ and $1^3D_2$ states, the states $1D(2^-)$ and $1D'(2^-)$ in the charmed-strange meson family satisfy the following relation
\begin{equation}
 \left(
  \begin{array}{c}
   |1D(2^-)\rangle\\
   |1D'(2^-)\rangle\\
  \end{array}
\right )=
\left(
  \begin{array}{cc}
    \cos\theta_{1D} & \sin\theta_{1D} \\
   -\sin\theta_{1D} & \cos\theta_{1D}\\
  \end{array}
\right)
\left(
  \begin{array}{c}
    |1^1D_2  \rangle \\
   |1^3D_2 \rangle\\
  \end{array}
\right),\label{1d}
\end{equation}
where $\theta_{1D}$ is a mixing angle. In the heavy quark limit, we can fix the mixing angle $\theta_{1D}=-50.8^\circ=-\arcsin(\sqrt{3/5})$ \cite{Godfrey:1986wj,Godfrey:2013aaa,Matsuki:2010zy}. Adopting the theoretical prediction of the masses of $D_s(1D(2^-))$ and $D_s(1D'(2^-))$ as input, we list their allowed decay channels in Table \ref{table:channel}.
The $\theta_{1D}$ dependence of the partial and total decay widths of $D_s(1D(2^-))$ and $D_s(1D'(2^-))$ is given in
Fig. \ref{fig:1d2}. When taking the limit value  $\theta_{1D}=-50.8^\circ $ \cite{Godfrey:1986wj,Godfrey:2013aaa,Matsuki:2010zy}, we conclude that the main decay modes of $D_s(1D(2^-))$ and $D_s(1D^\prime(2^-))$ are $D^\ast K$ and $DK^\ast$.  {In the heavy quark limit,  $D_s(1D(2^-))$ couples with $D^\ast K $ via a $P$ wave, while it couples with $D K^\ast $ only via a $F$ wave due to the conservation of light degree of freedom. As for $D_s(1D^\prime (2^-))$, the situation is opposite to $D_s(1D(2^-))$, $D_s(1D^\prime (2^-))$ strongly couples with $DK^\ast$ via a $P$ wave.  Our present calculations are consistent with the conclusion of the heavy quark limit \cite{Close:2005se, Burns:2014zfa} and the results are model independent in the heavy quark limit}. In addition, The total decay widths of $D_s(1D(2^-))$ and $D_s(1D^\prime(2^-))$ can reach up to 240 MeV and 147 MeV, respectively. The above study also shows that the $D^\ast K$ and $DK^\ast$ modes are the key channels when distinguishing the $D_s(1D(2^-))$ and $D_s(1D^\prime(2^-))$ states.

Besides $D_{s1}^*(2860)$, the LHCb collaboration also observed $D_{s3}^\ast(2860)$. In this work, we test whether $D_{s3}^\ast(2860)$ can be a good candidate of $D_s(1^3D_3)$. The partial and total decay widths of $D_{s3}^\ast(2860)$ as $D_s(1^3D_3)$ are listed in Table \ref{table:channel}, in which we find that our calculated total width is about one-fourth of experimental data \cite{Aaij:2014xza,Aaij:2014baa}.
The main decay modes of $D_s(1^3D_3)$ are $DK$ and $D^*K$, where the ratio
\begin{eqnarray}
\frac{\mathcal{B}(D_s(1^3D_3)\rightarrow D^{*0}K)}{\mathcal{B}(D_s(1^3D_3)\rightarrow D^0K)}=0.802
\end{eqnarray}
is obtained. We suggest to do a more precise measurement of the resonance parameters and the
ratio ${\mathcal{B}(D^{*0}K)}/{\mathcal{B}(D^0K)}$  of $D_{s3}^\ast(2860)$, which will finally make a definite conclusion of whether $D_{s3}^\ast(2860)$ is a $D_s(1^3D_3)$ state.

\subsubsection{$2P$ states}

There are four $2P$ states. Among these, $D_s(2^3P_0)$ and $D_s(2^3P_2)$ are still missing in experiment, while there exist possible candidates for $D_s(2P(1^+))$ and $D_s(2P^\prime(1^+))$.

The partial and total decay widths of $D_s(2^3P_0)$ are shown in Table \ref{table:channel}, where the mass of $D_s(2^3P_0)$ is fixed as

2960 MeV (see Table. \ref{table:spetrum}). The total decay width of$D_s(2^3P_0)$ is 166.15 MeV and there exist three main decay modes, $DK$, $D^*K^*$, and $D_{1}(2420)K$ (see Table. \ref{table:decay}).

As for the $D_s(2^3P_2)$ state with the predicted mass $3004$ MeV, the main decay modes are $D^*K^*$, $D^*K$, and $D_1(2430)K$ with the total decay width 86.25 MeV. The ratio is predicted as
$\mathcal{B}(D_s(2^3P_2)\rightarrow D^\ast
K)/\mathcal{B}(D_s(2^3P_2)\rightarrow DK^*)=0.39$.

We notice that there is an evidence of the structure around 2960 MeV in the $\bar{D}^0 K^-$ invariant mass spectrum given by LHCb \cite{Aaij:2014xza,Aaij:2014baa} except for the observed $D_{s1}^*(2860)$ and $D_{s3}^*(2860)$. If this evidence will be confirmed by future experiment, this structure around 2960 MeV must be either
$D_s(2^3P_0)$ or $D_s(2^3P_2)$ since there are no other candidates for natural states from 2800MeV to 3100 MeV as shown in Fig. \ref{fig:spectrum}. According to the calculated decay behaviors of $D_s(2^3P_0)$ and $D_s(2^3P_2)$, we can exclude  the possibility of the
$D_s(2^3P_2)$ assignment to this structure since the decay width of
$D_s(2^3P_2)\to DK$ is quite small.

In the following, we discuss the possibilities of the observed $D_{sJ}^*(3040)$ as $D_s(2P(1^+))$ or $D_s(2P^\prime(1^+))$, where
$D_s(2P(1^+))$ or $D_s(2P^\prime(1^+))$ satisfy
\begin{equation}
 \left(
  \begin{array}{c}
   |2P(1^+)\rangle\\
   |2P^\prime(1^+)\rangle\\
  \end{array}
\right )=
\left(
  \begin{array}{cc}
    \cos\theta_{2P} & \sin\theta_{2P} \\
   -\sin\theta_{2P} & \cos\theta_{2P}\\
  \end{array}
\right)
\left(
  \begin{array}{c}
    |2^1P_1  \rangle \\
   |2^3P_1 \rangle\\
  \end{array}
\right)\label{m4}
\end{equation}
with a mixing angle $\theta_{2P}$.  If taking $\theta_{2P}=\theta_{1P}=-54.7^\circ$ \cite{Godfrey:1986wj,Matsuki:2010zy} to estimate the decay behaviors of
$D_s(2P(1^+))$ or $D_s(2P^\prime(1^+))$, we obtain the numerical results listed in Table \ref{table:channel}.

If assigning $D_{sJ}^*(3040)$ to $D_s(2P(1^+))$, the calculated total decay width is 285.83 MeV, which is consistent with the experimental width \cite{Aubert:2009ah}. The main decay modes are $D^*K$, $DK^*$, $D^*K^*$ and $D_{2}^*(2460)K$, which can explain why $D_{sJ}^*(3040)$ was first observed by experiment in the $D^*K$ channel \cite{Aubert:2009ah}. Additionally, we predict the ratio
\begin{eqnarray}
\frac{\mathcal{B}(D_s(2P(1^+))\rightarrow D^\ast
K)}{\mathcal{B}(D_s(2P(1^+))\rightarrow DK^\ast)}=1.59,\label{jjj}
\end{eqnarray}
which can be tested by future experiment. The above study shows that $D_{sJ}^*(3040)$ is a good candidate of $D_s(2P(1^+))$, which is the first radial excitation of $D_{s1}(2460)$.

We need to discuss another possible assignment of $D_{sJ}^*(3040)$ to $D(2P^\prime(1^+))$ if considering study of only the mass spectrum shown in Fig. \ref{fig:spectrum} because the mass of $D_{sJ}^*(3040)$ is very close to the predicted mass of $D_s(2P^\prime(1^+))$. Moreover, the total decay width of $D_s(2P^\prime(1^+))$ is 131.28 MeV, which is comparable to the lower limit of experimental width of $D_{sJ}^*(3040)$ \cite{Aubert:2009ah}. The predicted ratio is, however,
\begin{eqnarray}
\frac{\mathcal{B}(D_s(2P^\prime(1^+))\rightarrow D^\ast
K)}{\mathcal{B}(D_s(2P^\prime(1^+))\rightarrow DK^\ast)}=3.19,
\end{eqnarray}
which is quite different from that given in Eq. (\ref{jjj}). Thus, we suggest to carry out the measurement of the ratio $\mathcal{B}(D_{sJ}^*(3040)\rightarrow D^\ast
K)/\mathcal{B}(D_{sJ}^*(3040)\rightarrow DK^\ast)$, which will be helpful to finally identify the inner structure of $D_{sJ}^*(3040)$.

\subsubsection{$3S$ states}

In this subsection, we predict the decay behaviors of two $3S$ states in the charmed-strange meson family, which are still missing in experiment. As shown in Table \ref{table:channel}, the partial and total decay widths of $D_s(3^1S_0)$ and $D_s(3^3S_1)$ are calculated.

As for $D_s(3^1S_0)$, the total decay width is about 111.98 MeV and
the main decay channels are $D^*K$, $D^*K^*$ and $D_0^*(2400)K$. As for $D_s(3^3S_1)$, the total decay width is 115.35 MeV and the main decay channels are $DK$, $D^*K$, $DK^*$ and $D_1(2430)K$. These informations are valuable for experimental search for these two missing  $D_s(3^1S_0)$ and $D_s(3^3S_1)$ mesons.

As the unnatural state, the predicted mass (3092 MeV) of $D_s(3^1S_0)$ is very close to the mass of $D_{sJ}^*(3040)$ reported by the Babar Collaboration \cite{Aubert:2009ah}. Thus, we again check
whether $D_{sJ}^*(3040)$ is explained as $D_s(3^1S_0)$. Since $D_{sJ}^*(3040)$ was observed in the $D^*K$ channel which is one of the main decay channels of $D_s(3^1S_0)$, this fact cannot exclude this possibility. However, the obtained total decay width of $D_s(3^1S_0)$ is a little bit smaller than
the lower limit of the experimental width of $D_{sJ}^*(3040)$. Although the predicted $D_s(3^1S_0)$ can not fit all the experimental feature of $D_{sJ}^*(3040)$, there exists small possibility of $D_{sJ}^*(3040)$ as $D_s(3^1S_0)$ if considering large experimental error of $D_{sJ}^*(3040)$. Thus, a crucial task in future experiment is the precise measurement of the resonance parameters of $D_{sJ}^*(3040)$, which can help us to have a definite conclusion of the properties of $D_{sJ}^*(3040)$.

\subsubsection{$2D$ states}

There are four $2D$ states all of which are absent in experiment. In Table \ref{table:2d}, we predict their decay properties. As for the $D_s(2^3D_1)$ state, the predicted mass by the modified GI model is 3244 MeV and the total width obtained in the QPC model is 165.20 MeV. Its main decay channels are $DK$, $D^*K$ and $D_1(2420)K$ and the typical decay ratio is given by
\begin{eqnarray}
\frac{\mathcal{B}(D_s(2^3D_1))\rightarrow D^\ast
K)}{\mathcal{B}(D_s(2^3D_1))\rightarrow DK)}=0.31.\label{23d1}
\end{eqnarray}

As an admixture of $2^1D_2$ and $2^3D_2$ states, $D_s(2D(2^-))$ and  $D_s(2D'(2^-))$ satisfy the following relation
\begin{equation}
 \left(
  \begin{array}{c}
   |2D(2^-)\rangle\\
   |2D'(2^-)\rangle\\
  \end{array}
\right )=
\left(
  \begin{array}{cc}
    \cos\theta_{2D} & \sin\theta_{2D} \\
   -\sin\theta_{2D} & \cos\theta_{2D}\\
  \end{array}
\right)
\left(
  \begin{array}{c}
    |2^1D_2  \rangle \\
   |2^3D_2 \rangle\\
  \end{array}
\right),\label{2d}
\end{equation}
where the mixing angle can be fixed as $\theta_{2D}=-50.8^\circ$ in the heavy quark limit \cite{Godfrey:1986wj,Godfrey:2013aaa,Matsuki:2010zy} when discussing their decay behaviors.

The $D_s(2D(2^-))$ with the predicted mass 3238 MeV has the total width 141.49 MeV and its main decay channels are $D^*K$ and $D^*_{s2}(2460)K$ that contribute to almost 80\% of the total decay width.
As the partner of $D_s(2D(2^-))$, $D_s(2D^\prime(2^-))$ has the predicted mass 3260 MeV, is a narrow state compared with $D_s(2D(2^-))$ because the total decay widths is 52.4 MeV, and its main decay mode is $DK^*$ with the partial decay width 30.67 MeV. The difference of the total decay widths of $D_s(2D(2^-))$ and $D_s(2D^\prime(2^-))$ can be understood since the decays of $D_s(2D(2^-))$ and $D_s(2D^\prime(2^-))$ into $D^*K$ occur via $P$-wave and $F$-wave in the heavy quark limit \cite{Godfrey:2013aaa}, respectively.

As for $D_s(2^3D_3)$ with mass 3251 MeV, the total decay width is 44.85 MeV and the main decay channels are $D^*K^*$ and $D_1(2430)K$. Here, the ratio
\begin{eqnarray}
\frac{\mathcal{B}(D_s(2^3D_3))\rightarrow D^*K)}{\mathcal{B}(D_s(2^3D_3))\rightarrow DK)}=0.18\label{23d3}
\end{eqnarray}
is predicted via the QPC model.

\renewcommand{\arraystretch}{1.5}
\begin{table}[htbp]
\caption{Decay behaviors of four $2D$ charmed-strange mesons (in the unit of MeV). }
\centering
\begin{tabular}{c c c c c }\toprule[1pt]
Channels&$D_s(2^3D_1)$&$D_s(2D(2^-))$&$D_s(2D^\prime(2^-))$ &$D_s(2^3D_3)$\\ \midrule[1pt]
$DK$&                                 65.02&          --&        --&5.25         \\
$D_s\eta$&                            4.13 &          --&        --&0.23            \\

$D_s\eta'$&                            0.87&          --&        --&0.02             \\
$D^{*}K$&                             20.32&       76.78&        0.03&0.96           \\
$D_{s}^*\eta$&                         0.93&       4.12 &        0.006&0.04           \\
$D^*_s\eta'$&                         0.04&        0.29&        0.002&$3.0\times10^{-7}$          \\
$DK^*$&                               5.03&       4.73&        30.67&0.69            \\
$D_s\phi$&                            0.06&        0.12&          0.04& 0.04            \\
$D^*K^*$&               $3.2\times10^{-4}$&        1.31&        3.11& 19.64             \\
$D_s^*\phi$&                          0.37&       0.42 &        0.47&0.36               \\
$D_0^*(2400)K$&                     --&        6.82&        8.00&--            \\
$D_{s0}^*(2317)\eta$&               --&       0.36 &        0.37&--             \\
$D_0^*(2400)K^*$&                 0.54&        0.16&        0.53&0.63            \\
$D_1(2430)K$&                       14.25&        8.14&        5.64&11.50                \\
$D_1(2420)K$&                      39.90&       2.80 &        1.71&1.26               \\
$D_{s1}(2460)\eta$&                 0.59&        0.32&        0.24&0.43               \\
$D_{s1}(2536)\eta$&                   0.19&        0.07&        0.06&0.007      \\
$D_2^*(2460)K$&                       12.77&       34.22&        1.46&3.74                \\
$D_{s2}^*(2573)\eta$&                 0.19&        0.83&        0.06&0.05               \\ \midrule[1pt]
Total&                              165.20 &     141.49 &     52.40  &44.85               \\ \bottomrule[1pt]
\end {tabular}\\
\label{table:2d}
\end{table}

\subsubsection{$1F$ states}

Similar to the experimental situation of the $2D$ states in the charmed-strange meson family, four $1F$ charmed-strange mesons are still missing in experiment. Thus, in the following we predict their decay behaviors.

The mass of $D_s(1^3F_2)$ is 3159 MeV predicted by the modified GI model. The OZI-allowed two-body strong decay channels are listed in Table \ref{table:1f}. Our calculation shows that its main decay modes are $DK$, $D^*K$, $DK^*$, and $D_1(2420)K$. The $D_s(1^3F_2)$ is a broad state with the total decay width 415.97 MeV, where the predicted ratio is
\begin{eqnarray}
\frac{\mathcal{B}(D_s(2^3F_2))\rightarrow D^\ast
K)}{\mathcal{B}(D_s(2^3F_2))\rightarrow DK)}=0.75.\label{23f2}
\end{eqnarray}

The $D_s(1F(3^+))$ and  $D_s(1F'(3^+))$ satisfy
\begin{equation}
 \left(
  \begin{array}{c}
   |1F(3^+)\rangle\\
   |1F'(3^+)\rangle\\
  \end{array}
\right )=
\left(
  \begin{array}{cc}
    \cos\theta_{1F} & \sin\theta_{1F} \\
   -\sin\theta_{1F} & \cos\theta_{1F}\\
  \end{array}
\right)
\left(
  \begin{array}{c}
    |1^1F_3  \rangle \\
   |1^3F_3 \rangle\\
  \end{array}
\right),\label{2f}
\end{equation}
where $\theta_{1F}$ is the mixing angle, which can be determined as $\theta_{1F}=-49.1^\circ=-\arcsin(2/\sqrt{7})$ in the heavy quark limit \cite{Godfrey:1986wj,Matsuki:2010zy}.
In Table \ref{table:1f}, we collect the calculated decay widths of $D_s(1F(3^+))$ and  $D_s(1F'(3^+))$, where we take
3139 MeV and 3169 MeV as their mass input, respectively. The main decay channels of $D_s(1F(3^+))$ are $D^*K$, $D^*K^*$ and, $D^*_2(2460)K$ and the total decay width can reach up to 372.48 MeV. Thus, $D_s(1F(3^+))$ has a very broad width. As for $D_s(1F'(3^+))$, $DK^*$ and $D^*K^*$ are its main decay modes and the total width of $D_s(1F'(3^+))$ is 193.34 MeV.

The $D_s(1^3F_4)$ with the mass 3143 MeV dominantly decays into $D^*K^*$ and the total decay width of $D_s(1^3F_4)$
is 150.79 MeV. In addition, we predict the following ratios
\begin{eqnarray}
\frac{\mathcal{B}(D_s(2^3F_4))\rightarrow D
K)}{\mathcal{B}(D_s(2^3F_4))\rightarrow D^\ast K)}=0.77,\label{23f4a}
\end{eqnarray}
and
\begin{eqnarray}
\frac{\mathcal{B}(D_s(2^3F_4))\rightarrow D
K^\ast)}{\mathcal{B}(D_s(2^3F_4))\rightarrow DK)}=0.82,\label{23f4b}
\end{eqnarray}
which can be tested in future experiment.

\renewcommand{\arraystretch}{1.5}
\begin{table}[htbp]
\caption{Decay behaviors of four $1F$ charmed-strange mesons ( in the unit of MeV). }
\centering
\begin{tabular}{c c c c c }\toprule[1pt]
Channels&$D_s(1^3F_2)$&$D_s(1F(3^+))$&$D_s(1F'(3^+))$ &$D_s(1^3F_4)$\\ \midrule[1pt]
$DK$&                                  57.82&          --&            --&5.30         \\
$D_s\eta$&                              3.16&          --&            --& 0.11             \\
$D_s\eta'$&                             0.41&          --&            --&0.002             \\
$D^{*}K$&                              43.12&       98.10&         14.86&6.88            \\
$D_{s}^*\eta$&                          2.19&        4.98&          0.28&0.11            \\
$D^*_s\eta'$&                           0.06&        0.08&$4.3\times10^{-4}$&$5.5\times10^{-5}$ \\
$DK^*$&                                41.12&        9.34&        102.32&4.35             \\
$D_s\phi$&                              0.85&        0.04&          2.36&0.08              \\
$D^*K^*$&                             16.68&       51.12&         66.16& 130.87              \\
$D_s^*\phi$&                           0.01&       0.002&         0.09 &0.01               \\
$D_0^*(2400)K$&                          --&        0.50&          3.32&--            \\
$D_{s0}^*(2317)\eta$&                    --&        0.02&          0.16&--             \\
$D_1(2430)K$&                          0.49&        0.22&          0.68&1.24                \\
$D_1(2420)K$&                        220.10&        1.40&          2.14&0.42                \\
$D_{s1}(2460)\eta$&                   0.007&        0.003&          0.02&0.02               \\
$D_{s1}(2536)\eta$&                  5.35&$6.8\times10^{-4}$&     0.004&$3.0\times10^{-4}$      \\
$D_2^*(2460)K$&                      24.34&       205.68&          0.95&1.40                 \\
$D_{s2}^*(2573)\eta$&                 0.26&         0.99&$7.9\times10^{-4}$&$6.5\times10^{-5}$ \\ \midrule[1pt]
Total&                              415.97 &       372.48&        193.34 &150.79        \\ \bottomrule[1pt]
\end {tabular}\\
\label{table:1f}
\end{table}

\section{Summary \label{sec5}}

In the paste decade, more and more charmed-strange states have been reported by different experiments, which are collected in Table \ref{table:review}. The present status of these observed charmed-strange states has stimulated us to revisit the charmed-strange meson family and to systemically carry out the study of their mass spectra and two-body OZI-allowed strong decays.

In this work, we have adopted the modified GI model to get the mass spectra of charmed-strange meson family, where the screening effect partly reflecting the unquenched effect is considered in our calculation. Comparing the theoretical results with the experimental data, we can roughly obtain the properties of the observed charmed-strange mesons. In addition, we have also predicted the masses of the higher radial and orbital excitations in the charmed-strange meson family.
This information is important in searching for these missing higher charmed-strange mesons in future experiment. Besides the mass information, in our calculation we have obtained the numerical results of the spatial wave functions of the discussed charmed-strange mesons, which are applied to calculate their two-body OZI-allowed strong decays.

To obtain the decay behaviors of the discussed charmed-strange mesons, we have adopted the QPC model in this work, where the $1P$, $2P$, $1D$, $2S$, $2D$, $3S$, and $1F$ states in the charmed-strange meson family are involved. The study of their strong decay behaviors further test the possible assignments to the observed states, where a phenomenological analysis is performed. As for the higher charmed-strange mesons absent in experiment, we have predicted their partial and total decay widths, and some typical decay ratios, which is valuable for experimental study of these states.

Since 2003, the BaBar, Belle, CLEOc, and LHCb experiments have made a big progress on the search for charmed-strange mesons. In the next ten years, we believe that more candidates of the charmed-strange mesons will be reported with running of the LHC experiment at 14 TeV collision energy, forthcoming BelleII and PANDA. The study presented in this work is helpful to identify these observed charmed-strange states and carry out a search for higher radial and orbital excitations in the charmed-strange meson family.

\section*{Acknowledgments}

This project is supported by the National Natural Science
Foundation of China under Grants No. 11222547, No. 11175073, No. 11375240, and No. 11035006, the Ministry of Education of China (SRFDP under Grant No. 2012021111000), and the Fok Ying Tung Education Foundation (No. 131006).

\noindent{\bf Note added}: {After submitting our paper to arXiv, we have noticed a very recent work in arXiv:1502.03827 \cite{Segovia:2015dia}, in which the authors also studied the mass spectrum and strong decays of charmed-strange mesons and obtained the results similar to the present work.}

\end{document}